\documentclass[preprint]{revtex4}
\usepackage{amssymb}
\usepackage{makeidx}
\usepackage{amsmath}
\usepackage{graphicx}
\usepackage{dcolumn}
\usepackage{bm}
\usepackage{epstopdf}

\begin{document}

\title{Relativistic Classical and Quantum Nonlinear Phenomena in the Induced
Processes on Free Electrons}
\author{H.K. Avetissian}
\affiliation{Centre of Strong Fields Physics, Yerevan State University, 1 A. Manukian,
Yerevan 0025, Armenia}
\date{\today }

\begin{abstract}
 With the appearance of superpower laser sources of
relativistic/ultrarelativistic intensities in the last decade, the
laser-QED-vacuum-matter interaction physics has entered a new phase that
makes real the observation of many nonlinear quantum electrodynamic (QED)
and classical-quantum- mechanical phenomena revealed more than fore-five
decades ago, serious advance in new generation of laser-plasma accelerators
of ultrahigh energies, nuclear fusion etc. Hence, the present review article
will help explorers-experimentalists in this field to attract attention on
the fundamental properties and peculiarities of the dynamics of induced
free-free transitions at high and superhigh intensities of stimulated
radiation fields. In this connection it is of special interest the induced
Cherenkov, Compton, undulator/wiggler coherent processes, as well as
cyclotron resonance in a medium -- possessing with nonlinear resonances of
threshold nature and leading to many important nonlinear effects or
applications, which are considered in this review. 
\end{abstract}

\maketitle

\tableofcontents



\section{Introduction}

The unprecedented development of laser technologies in recent two decades,
mainly due to chirped pulse amplification technique \cite{Chirp}, has
allowed increase of intensity of laser sources in the first stage a six
orders of magnitude \cite{Str_Mour,Mour_92,Mour_94,Esar,Mour_98} reaching
the relativistic intensities $10^{16}\mathrm{W/cm}^{2}$--$10^{18}\ \mathrm{%
W/cm}^{2}$ in the infrared and optical domains, respectively, then exceeding
these values up to $10^{22}\ \mathrm{W/cm}^{2}$ for current superpower laser
beams of ultrarelativistic intensities \cite{Danson,Mourou}. In many
laboratories \cite{Lab1,Lab2,Lab3,Lab4} compact lasers can deliver $1-10$
petawatt short pulses, with focused intensities as high as $10^{22}\ \mathrm{%
W\ cm}^{-2}$. Next generation multipetawatt and exawatt optical laser
systems \cite{ELI,XCELS,HiPER} will be capable to deliver ultrahigh
intensities exceeding $10^{25}\ \mathrm{W\ cm}^{-2}$, which are well above
the ultrarelativistic regime of electron-laser interaction. At such
intensities the radiation-matter interaction enters the phase of ultrafast
dynamics in supershort time scales, with formation of such important fields
in Condensed-Matter-Physics that the Attoscience \cite%
{At1,At2,Atto1,Atto2,Atto3,Atto4}, and Relativistic Optics \cite%
{Mourou,Rel-Rep2,Br.-Kr.,Rel-Rep1,Pukh} are. \ \ 

The appearance of laser sources with ultrarelativistic intensities has
opened real possibilities for revelation of many nonlinear electrodynamic
phenomena at the interaction of superpower radiation fields with high
brightness accelerator beams \cite{2006}, relativistic plasma targets of
solid densities \cite{Mey.-ter,Esir.,Ter-Av.,Yan,Naum.1,Naum.2}, and QED
vacuum \cite{2016}.\textrm{\ }Laser sources of such enormous intensities
predetermine, in particular, the future of laser accelerators
 \cite{18_exp,19_exp,20_exp,21_exp,22_exp,23_exp} of superperhigh energies, including
laser-plasma accelerators \cite{Mey.-ter,1p,2p,3p,4p,5p,6p,7p,8p}. Thus,
currently available optical lasers provide $\xi $ up to $100$, meanwhile
with the next generation of laser systems one can manipulate with beams at $%
\xi $ $>1000$. In such ultrastrong laser fields, electrons can reach to
ultrarelativistic energies. However to obtain ultrarelativistic net energies
for field-free electrons we need the third body for satisfaction of
conservation laws for energy-momentum of a free electron (after the
interaction) with absorption/radiation of photons in the field of a plane
monochromatic wave \cite{2006}, or take advantage of nonplane character of
tightly focused supershort laser pulses \cite{BNL_Accel}. Nevertheless, the
actual way to get net energy exchange of relativistic electrons is using of
coherent processes of laser-particle interaction with the additional
resonances. Among those the induced Cherenkov, Compton, and undulator
processes are especially of interest \cite{2006}. These are coherent induced
processes where a critical intensity of stimulated wave exists (because of
coherent character of corresponding spontaneous process) above which the
induced process proceeds only in one direction - coherent radiation (wave
amplification), or absorption (particle acceleration) depending on the
particle initial velocity with respect to resonance value (satisfying the
condition of coherency) of each process. The latter is the most important
feature of nonlinear resonance inherent in these processes to get high net
energy change of particles, specifically for laser acceleration \cite{2016},
or free electron laser (FEL) \cite{Brau,Freund}, as well as to generate
relativistic electron beams of low energy spreads and emittances due to the
threshold nature of nonlinear resonance in these induced coherent processes.

In principle, the realization of acceleration or inverse problem of FEL in
any laser-induced process requires the possible largest coherent length of
particle-laser stimulated interaction at which the accumulation of particle
energy in the acceleration regime, or coherent radiation -in the wave
amplification regime take place on the whole coherent length of certain
process. In other cases, the existence of an additional resonance (e.g., in
a static magnetic field \cite{autores,autores2,Hyst,Buchb,MCR,Quant_CR}) is
necessary where the particle-wave interaction cross section enhances by many
orders in magnitude. These are the linear resonances well enough
investigated during the past decades. Here we will consider laser-induced
processes on free electrons with the large effective gain of interaction due
to nonlinear resonance achieved in the given field of a strong laser
radiation. In the induced Cherenkov, Compton, and undulator processes such
resonance is adequate to fulfilment of condition of coherency between the
particle and wave achieved in the field at the aforementioned critical
intensity of the total wave-field. In the result of such interaction, at the
laser intensities larger than critical value a nonlinear threshold
phenomenon of particles \textquotedblleft reflection\textquotedblright\ or
capture take place -accelerating or decelerating electrons on the shortest
interaction lengths, even shorter than a laser wavelength \cite{Cher._Reflec,Comp._Reflec,Und._Reflec}.

In contrast to induced Cherenkov process \cite{Cher._Reflec} where the
intensities of applied laser fields are confined by ionization threshold of
dielectric media \cite{Keldish,Bunkin} in vacuum processes -- in the induced
Compton \cite{Comp._Reflec} and undulator \cite{Und._Reflec} ones there is
no restriction on the electromagnetic (EM) field strengths that allows to
apply current superpower laser beams of ultrarelativistic intensities \cite%
{Mourou} for acceleration of particles up to superhigh energies. The other
practical difference between these induced processes is in the relativistic
character of initial resonance width of particles to reach the threshold
value of corresponding nonlinear resonance in the field for
\textquotedblleft reflection\textquotedblright /capture phenomenon. Thus, in
contrast to Cherenkov and undulator processes where\ practically initial
relativistic/ultrarelativistic particles are needed for realization of
nonlinear Cherenkov resonance in gaseous media \cite{last}, or nonlinear
resonance in a magnetic undulator with the step of the order of a few $%
\mathrm{cm}$ \cite{wigler}, induced Compton mechanism for particles
\textquotedblleft reflection\textquotedblright /capture and, consequently,
acceleration regime practically may be realized for arbitrary initial
energies of the particles \cite{2_beams}. In particular, using laser pulses
of near frequencies one can accelerate initially nonrelativistic beams, or
even particles in rest. Second, at the same frequencies of
counterpropagating laser pulses in the induced Compton process one can
achieve the cancellation of particles transverse momenta acquired in the
field to obtain quasi-colinear particles bunches from the emerging
relativistic electron targets (elimination of a bunch angular divergence),
as it has been proposed in the paper \cite{Meyer-ter-Vehn} for generation of
uniform relativistic electron layers at the coherent Compton backscattering
on the ultra-thin solid foils (see, also, \cite{other,other2}).

A promising way for achieving of laser-driven electron acceleration is the
use of a plasma medium \cite{Plasma}. However, the laser-plasma accelerator
schemes face problems connected with the inherent instabilities in
laser-plasma interaction processes. On the other hand, the spectrum of
direct acceleration mechanisms of charged particles by a single laser pulse
is very restricted, since one should use the laser beams focused to
subwavelength waist radii, or use subcycle laser pulses, or use radially
polarized lasers, violating the plane character of a coherent laser pulse
for particles acceleration by a single wave-pulse in vacuum without the
third body. All these scenarios with different field configurations have
been investigated both theoretically and experimentally in the works \cite%
{22_exp,A1,A2,A3,A4,A5,A6,A7,A8,A9,A10,A11,A12,A13,A14,A15,A16,A17}. Note
that beside the focusing of a laser pulse to subwavelength waist radii we
can get nonplane laser beam by terminating the field, either by reflection,
absorption, or diffraction \cite{Pant}. The proof of principle experiment of
this type has been reported in Ref. \cite{PPP}. Here it is used initially
relativistic electron beam and a moderately strong laser pulse. To obtain
dense enough electron bunches it is reasonable to consider electrons
acceleration from nanoscale-solid plasma-targets \cite%
{Meyer-ter-Vehn,mirror1,mirror2,mirror3} at intensities high enough to
separate all electrons from ions \cite{Kulagin}. Thus, combining these two
schemes one can obtain ultrarelativistic solid density electron bunches \cite%
{11b,12b,13b,14b,15b,16b,17b,18b}. Such bunches can be used to obtain
high-flux of positrons, $\gamma $-quanta \cite{mer_rel._tirax1,mer_rel._tirax2}
with possible applications in material science, medicine, and
nuclear physics.

The recent achievements in the laser technology related to implementation of
supershort--femtosecond laser sources and subsycle pulses of relativistic
intensities exceeding the intra-atomic fields open real opportunities for
realization of laser-plasma accelerators of ultrahigh energies \cite%
{Mey.-ter,1p,2p,3p,4p,5p,6p,7p,8p}. These schemes are based on the
laser-plasma-wake-field and laser-beat-wave mechanisms of laser-assisted
acceleration in plasma. The interaction of such powerful radiation fields
with the plasma has led to generation of quasi-monoenergetic relativistic
electron beams in laser-driven plasma accelerators \cite%
{A9,A10,A11,A12,A13,A14,A15,A16,A17}. Concerning the interaction of
ultrashort laser pulses of relativistic intensities with the free electrons
in vacuum, i.e. the implementation of laser accelerators, one should note
that experiments in this area are gathering power at present.

One of the first laser accelerator concepts is the Inverse Cherenkov
Accelerator (ICA). The ICA scheme has been investigated since the advent of
lasers [see, for example, \cite{Shim}]. The appearance of powerful laser
sources already in the next decade after the first lasers has stimulated
comprehensive theoretical \cite{Cher._Reflec,capture,Ch._K-G,spin,izv,Dekker,dif.3,ushir,mod,dis,Suplum} and experimental \cite%
{P1,P2,P3,P4,P5,Kim1,Kim2,Kim3,Kim4} investigations towards the
implementation of both induced Cherenkov problems --ICA and Cherenkov laser
realization-- in diverse schemes of multiphoton interaction that makes a
phase velocity matching of EM wave with a particle by controlling the
refraction index of gaseous medium so as to be a wave phase velocity in a
dielectric medium less than light speed in vacuum. Various interaction
schemes of electrons with stimulating radiation have been considered as with
ultrarelativistic electron beams in a gaseous medium for the optical region 
\cite{capture,Dekker,Suplum,izv,mod}, as well as with mildly relativistic
beams in dielectric waveguides for the microwave region \cite%
{W1,W2,W3,W4,W5,W6,W7,W8,W9,unmag}. In the latter case, the electron beam
passes over the dielectric in the vacuum --surface Cherenkov process-- that
enables one to avoid the impeding factors of the medium (the multiple
scattering of electrons on atoms, the ionization losses in the medium etc.).
Besides, the usage of solid-state dielectric as a slow-wave structure (with
rather high dielectric constant $\varepsilon $, $\varepsilon -1\sim 1$)
enables the achievement of electron-wave synchronism by mildly or moderately
relativistic electron beams.

It is important to note that the effect of critical field exists even in a
very week wave field if the interaction occurs in the region close to the
initial Cherenkov resonance. Therefore one can not escape these
peculiarities restricting only the intensity of the external wave field and
solve the problem in the scope of the linear theory even for very week wave
fields \cite{UFN}. Note that regarding the induced Cherenkov process a
series of systematical wrong works have been made during the last two
decades of past century, which were discussed in special review \cite{UFN}
devoted to consideration of those results. In the mentioned review the all
principal mistakes have been shown in detail, therefore in the current
review article we will not repeat the analysis of those papers which reader
can get in the review \cite{UFN}. The particle \textquotedblleft
reflection\textquotedblright\ phenomenon from a slowed plane wave pulse has
been applied also for cases where the group velocity of a wave pulse due to
the dispersion can be less than light speed in vacuum \cite{34,35,36,37}: in
Ref. \cite{34,35} --for the case of a plasma, in Ref. \cite{36,37} --at the
focusing of an ultrashort laser pulse in vacuum, in the so-called by authors
\textquotedblleft dispersion-dominated propagation regime\textquotedblright .

Nonlinear resonance of threshold nature requiring shortest acceleration
length also takes place at the charged particle interaction with strong
laser radiation\textit{\ }and static magnetic field along the wave
propagation direction\textit{\ }(when the resonant effect of the wave on the
particle rotational motion in the uniform magnetic field is possible) in a
medium \cite{Hyst}. In vacuum, as a result of the interaction of a charged
particle with a monochromatic EM wave and uniform magnetic field the
resonance created at the initial moment for the free-particle velocity
automatically holds throughout the whole interaction process due to the
equal Doppler shifts of the Larmor and wave frequencies in the field. This
phenomenon is known as \textquotedblleft Autoresonance\textquotedblright\ 
\cite{autores,autores2}.\textit{\ }This property of cyclotron resonance in
vacuum makes possible as particles acceleration \cite{S-F-K}, as well as the
creation of a generator of coherent microwave radiation/radioemission by an
electron beam, namely, a cyclotron resonance maser (CRM) \cite%
{Ginz1,Ginz2,McN,Coo,Nus,Ai1,Ai2,Brat}.

From the point of view of quantum theory the relativistic nonequidistant
Landau levels of the particle in the wave field become equidistant in the
autoresonance with the quantum recoil at the absorption/emission of photons
by the particle.\textit{\ }In addition, the dynamic Stark effect of the wave
electric field on the transversal bound states of the particle does not
violate the equidistance of Landau levels in the autoresonance.\textit{\ }%
Then the inverse process, that is, multiphoton resonant excitation of Landau
levels by strong EM wave and, consequently, the particle acceleration in
vacuum due to cyclotron resonance, in principle, is possible.\textit{\ }

In a medium with arbitrary refraction properties (dielectric or plasma)
because of the different Doppler shifts of the Larmor and wave frequencies
in the interaction process the autoresonance is violated. However, the
threshold (by the wave intensity) phenomenon of electron hysteresis in a
medium due to the nonlinear cyclotron resonance in the field of strong
monochromatic EM wave takes place \cite{Hyst} (this process has been treated
in the earlier paper \cite{Buchb}, where, however, only an oscillating
solution has been obtained).\textit{\ }In contrast to autoresonance, the
nonlinear cyclotron resonance in a medium proceeds with a large enough
resonant width.\textit{\ }This so-called phenomenon of electron hysteresis
leads to significant acceleration of particles, especially in the plasmalike
media where the superstrong laser fields of ultrarelativistic intensities
can be applied \cite{2016}. Note that the use of dielectriclike (gaseous)
media in this process makes it possible to realize cyclotron resonance in
the optical domain (with laser radiation) due to the possibility of
arbitrarily small Doppler shift of a wave frequency close to the Cherenkov
cone, in contrast to the vacuum case where the\ cyclotron resonance for the
existing maximal powerful static magnetic fields is possible only in the
microwave domain \cite{autores,autores2}.

The particle \textquotedblleft reflection\textquotedblright\ phenomenon has
a significant property: at the above-threshold interaction the induced
process occurs only in one direction--either direct (laser scheme) or
inverse (accelerator scheme) \cite{2006,2016} meanwhile at the
below-threshold interaction the both direct and inverse processes take place
simultaneously that reduces in general the net gain for each problem.
Therefore this phenomenon gives in principal new opportunity for
implementation of nonlinear schemes of FEL or laser accelerators in the
induced coherent processes at the above-threshold interaction regime. Due to
this feature the \textquotedblleft reflection\textquotedblright\ phenomenon
may also be used for the monochromatization of particle beams \cite{Ch._mon.,Comp._mon.,und._mon.,conv.}. On the other hand, because of
\textquotedblleft reflection\textquotedblright\ phenomenon the nonlinear
Compton effect in media with $\varepsilon >1$ can proceed only at the wave$\ 
$intensities below the critical one \cite{izv}.

For the nonplane EM wave pulses of small space sizes (tightly focused) and
of short duration, such as current superpower laser pulses are, it is
impossible to solve the problem analytically. Hence, in the papers \cite%
{last,wigler,2_beams} the \textquotedblleft reflection\textquotedblright\
and capture phenomena in case of actual nonplane-short laser pulses has been
investigated by numerical integration of relativistic equations of motion.

The phenomenon of particle \textquotedblleft reflection\textquotedblright\
and capture by EM wave being of pure classical nature, nevertheless leads to
quantum effects of probability density modulation of a free particle state
on the hard x-ray frequencies because of interference of incident and
reflected (inelastic) de Broglie waves after the reflection from a wave
barrier \cite{Ch._K-G} and of zone structure of particle states in the wave
potential well --capture regime like the particle states in a crystal
lattice \cite{Q1,Q2,Q3}. On the other hand, it is evident that the role of
particle spin in this process\textit{\ }is important since in dielectriclike
media the wave periodic EM field in the intrinsic frame of reference becomes
a static magnetic field and spin interaction\textit{\ }with such a field
should resemble the Zeeman effect. Besides, the critical intensity depends
on the particle spin projection and, consequently, it is differ for
different initial spin orientations. Choosing the certain value for critical
intensity corresponding to the certain spin orientation of particles in the
beam, with the appropriate laser intensity (above this critical value) one
can achieve the particle beam polarization.These are the quantum effects in
the above critical regime of induced Cherenkov process. Below the critical
intensity the coherent effects of particles classical bunching on optical
and shorter wavelengths \cite{bunch.1,P1,P2,P3,Kim1}, quantum effects of
modulation of particle probability density on the wave harmonics \cite%
{mod,mod.2,mod.3}, and inelastic diffraction scattering of the particles on
the traveling EM wave in a dielectric medium takes place \cite%
{dif.2,dif.3,dif.4} like the Kapitza--Dirac effect on a standing wave
lattice \cite{dif.1} or Bragg diffraction on a crystal lattice \cite{1913}.
\ 

The interest to Kapitza--Dirac effect \cite{dif.1} has been increased
especially after the successful realization of experiment with
high-intensity and strongly coherent laser beams-gratings \cite{dif._exp.}.
It follows to note that observation of diffraction effect during the next 70
years after its prediction had not been succeeded. For the acquaintance with
the experimental situation in this area since the advent of this effect, or
for detailed references on earlier work we refer the reader to review papers 
\cite{Schw.1,Schw.2,Tak.,Buch,Bat}. The significance of Kapitza--Dirac
effect, apart from its quantum-mechanical meaning as a best example of
demonstration of electron matter wave diffracted by light and, moreover, as
an unique sample of a diffraction system with reversed properties of the
matter and light, is also conditioned by important applications, since
electron beams difracted from highly coherent laser gratings are coherent
with each other. Hence, the Kapitza--Dirac effect is a very convenient, even
maybe an irreplaceable means to realize coherent electron beams. Such beams
can serve as a basis for construction of new important tools of diverse
species, e.g., coherent beam splitters, new type electron interferometers
which would operate at rather low electron energies (typical for existing
now in atomic physics) etc. \ \ \ \ \ \ \ \ \ \ \ \ \ \ \ \ \ \ \ \ 

Kapitza--Dirac effect on standing wave lattice is a particular case of the
induced Compton process in the field of two counterpropagating EM\ waves of
the same frequencies at which electron moves in perpendicular direction to
wavevectors of counterpropagating waves (classical condition of resonance),
to exclude the Doppler shifting of waves' frequencies because of
longitudinal component of electron velocity. At the quantum condition of
resonance -taking into account the quantum recoil as well- the exact
condition of resonance is satisfied at the small angle to perpendicular
direction that corresponds to elastic Bragg diffraction on the phase lattice
of a standing wave \textrm{\cite{Bragg}}. Nevertheless, the phase matching
between the electrons and counterpropagating waves in the induced Compton
process can also be fulfilled in general case of bichromatic EM waves of
different frequencies if electron moves at the certain angle with respect to
wavevectors of counterpropagating waves at which the condition of coherency
between the electron and waves of different frequencies is satisfied \cite%
{dif.2}. However, in contrast to Kapitza--Dirac effect, diffraction of
electrons in this case is inelastic. Thus, due to the induced Compton effect
in the two wave fields an electron absorbs $s$ photons from the one wave and
coherently radiates $s$ photons into the other wave of different frequency
and vice versa. This is the condition of coherency in the induced Compton
process corresponding to the resonance between the Doppler-shifted
frequencies in the intrinsic frame of reference of an electron in the
bichromatic counterpropagating waves at which the conservation of the number
of photons in the induced Compton process takes place \cite%
{Fed.,dif.2,Fed.McI} (see, also the paper \cite{Fedorov} where theoretical
analysis of the scattering of electrons by a strong standing wave with a
slowly varying amplitude has been made), in contrast to spontaneous Compton
effect in the strong wave field where after the multiphoton absorption a
single photon is emitted \cite{Gold.}

So, the scope of the Kapitza--Dirac effect has been extended since 1975 in
the works \cite{dif.2,dif.3,dif.4} for inelastic diffraction scattering of
electrons on strong travelling wave in the induced Compton, Cherenkov, and
undulator/wiggler processes, taking into account the the mentioned
peculiarity in these processes. Note that the term "strong wave" here is
relative, since the mentioned nonlinear resonance takes place even in the
very weak wave-fields if the electrons initially are close to the resonance
state, i.e. the electrons' initial longitudinal velocity is close to the
phase velocity of the slowed wave, at which the critical field is also very
small and, respectively, very low wave intensities may be above critical 
\cite{UFN}. At this background, the paper \cite{mis} is completely
misunderstanding where authors, citing the papers \cite{dif.2,dif.3,dif.4}
but bypassing the existence of critical field with aforementioned
peculiarity, they obtain the same formula of the paper \cite{dif.3} for
multiphoton probability of diffraction scattering on the travelling wave in
a dielectric medium (by other way -on the base of the Helmholtz--Kirchhoff
diffraction theory- following to papers \cite{Echl-Lub,Echl-Lub2}), and
after four decades claim about the possibility of diffraction effect of
electrons on the travelling wave in a dielectric medium.

The organization of the paper is as follows. In Sec. II the exact nonlinear
theory of induced Cherenkov process in the strong laser fields is presented.
The peculiarity inherent to nonlinear resonance in coherent processes of
threshold character and existence of critical intensity leading to particle
\textquotedblleft reflection\textquotedblright\ and capture\ phenomenon is
described in Subsec. II 1. In Subsec. II 2. laser acceleration of charged
particles in gaseous media -- Cherenkov accelerator with the variable
refraction index of a gaseous medium is considered. In Subsec. II 3. the
nonlinear cyclotron resonance of threshold nature in a medium -- electron
hysteresis phenomenon is presented. In Subsec. II 4. the theory of coherent
radiation of charged particle beams in capture regime -- Cherenkov amplifier
is presented.

In Sec. III. the quantum theory of nonlinear induced Cherenkov process is
considered. In Subsec. III 2. the spin-effects at the electron
\textquotedblleft reflection\textquotedblright\ phenomenon and the
possibility to achieve the polarized particle beams is considered. In
Subsec. III 3. quantum coherent effect of electrons reflection from the
traveling-wave-phase-lattice in the exact resonance is investigated. In
Subsec. III 4. the quantum regime of spinor particles capture by slowed
traveling wave at the exact Bragg resonance is considered. In Subsec. III 5.
the quantum modulation effect of particle state on the optical harmonics at
the pump wave intensities below the critical value is presented.

In Sec. IV the vacuum versions of induced coherent processes -- induced
Compton and undulator/wiggler processes are investigated revealing the
particle \textquotedblleft reflection\textquotedblright\ nonlinear
threshold\ phenomena in vacuum, as well as inelastic diffraction effect of
electrons like to Kapitza--Dirac effect in vacuum are considered. In Subsec.
IV 1. induced Compton process in the field of two bichromatic
counterpropagating waves is investigated. In Subsec. IV 2. the
\textquotedblleft reflection\textquotedblright\ phenomenon in the
undulator/wiggler is considered. In Subsec. IV 3. the general theory of
inelastic diffraction effect of electrons on the slowing travelling wave is
presented. In Subsec. IV 4. diffraction regime of electron coherent
scattering on the traveling wave phase-lattice is considered. In Subsec. IV
5. the Bragg regime of exact resonance on travelling wave lattice is
presented. Finally, conclusions are given in Sec. V.

\section{Induced Cherenkov Process with Strong Laser Fields}

\subsection{Critical Intensity in the Induced Cherenkov Process. Nonlinear
Threshold Phenomenon of Particle \textquotedblleft
Reflection\textquotedblright\ and Capture}

Relativistic classical equations of motion for a charged particle energy in
the field of a plane EM wave give the exact solution \cite{Cher._Reflec}:

\begin{equation*}
\mathcal{E}\left( \tau \right) =\frac{\mathcal{E}_{0}}{n^{2}-1}\Biggl\{%
n^{2}\left( 1-\frac{\mathrm{v}_{0}}{cn}\cos \vartheta \right) \mp \biggl[%
\left( 1-n\frac{\mathrm{v}_{0}}{c}\cos \vartheta \right) ^{2}-\left(
n^{2}-1\right)
\end{equation*}%
\begin{equation}
\times \left( \frac{mc^{2}}{\mathcal{E}_{0}}\right) ^{2}\Bigl[\xi ^{2}\left(
\tau \right) \cos ^{2}\omega \tau -2\frac{p_{0}\sin \vartheta }{mc}\xi
\left( \tau \right) \cos \omega \tau \Bigr]\biggr]^{1/2}\Biggr\}  \label{1}
\end{equation}%
where $n=\sqrt{\varepsilon \mu }$ is the refraction index of a medium ($%
\varepsilon $ and $\mu $ are the dielectric and magnetic permittivities of
the medium, respectively), $m$, $\mathcal{E}_{0}$, $p_{0},\mathrm{v}_{0}$
are respectively particle mass, initial energy, momentum and velocity
-directed at the angle $\vartheta $ to the wave propagation direction $%
\mathbf{\nu }$ ($\left\vert \mathbf{\nu }\right\vert =1$), $c$ is the light
speed in vacuum, and $\xi \left( \tau \right) $ denotes the dimensionless
relativistic invariant intensity parameter of the wave: $\xi \left( \tau
\right) \equiv eA(\tau )/mc^{2}=eE(\tau )/mc\omega $ with the vector
potential $\mathbf{A}(t,\mathbf{r})=\mathbf{A}(t-n\mathbf{\nu r}/c)$; $\tau
=t-n\mathbf{\nu r}/c$ is the retarding wave coordinate of the plane EM wave
in a medium, and $A(\tau )$, $E(\tau )$ -- slowly varying amplitudes of the
vector potential and electric field of a quasimonochromatic wave amplitudes
(the wave is linearly polarized along the axis $OY$: $A_{y}=A(\tau )\cos
\omega \tau $).

Hereafter we will assume that the wave frequency $\omega $ is far from the
main resonance transitions between the atomic levels of the medium to
prohibit the wave absorption and nonlinear optical effects in the medium and
consequently $n=n(\omega )$ will correspond to the linear refraction index
of the medium.

As is seen from Eq. (\ref{1}) the expression determining the particle energy
in the wave field is, first, not single-valued and, second, may become
imaginary depending on particle and wave parameters. The peculiarity arising
in the induced Cherenkov process because of particle--strong wave nonlinear
interaction is connected with this fact. Hence, treatment of the particle
dynamics in this process should start by clarification of these questions.

To consider the behavior of a particle upon nonlinear interaction with a
strong wave in a medium on the basis of Eq. (\ref{1}) we will analyze the
case where the initial velocity of the particle at $\tau =-\infty $ ($%
\mathbf{A}(\tau )\mid _{\tau =-\infty }=0$) is directed along the wave
propagation direction for which the picture of the particle nonlinear
dynamics is physically more evident. In this case Eq. (\ref{1}) becomes: 
\begin{equation}
\mathcal{E}=\frac{\mathcal{E}_{0}}{n^{2}-1}\left[ n^{2}\left( 1-\frac{%
\mathrm{v}_{0}}{cn}\right) \mp \left( 1-n\frac{\mathrm{v}_{0}}{c}\right) 
\sqrt{1-\frac{\xi ^{2}\left( \tau \right) }{\xi _{cr}^{2}}}\right] ,
\label{5}
\end{equation}%
\begin{equation}
\xi _{cr}\equiv \frac{\mathcal{E}_{0}}{mc^{2}}\frac{|1-n\frac{\mathrm{v}_{0}%
}{c}|}{\sqrt{n^{2}-1}}.  \label{6}
\end{equation}

As is seen, Eq. (\ref{5}) is twovalence and, at first, we shall provide the
unique definition of the particle energy in accordance with the initial
condition. In the case of plasma ($n<1$ ) or vacuum ( $n=1$ ) the term under
the root is always positive, hence, in these cases one has to take before
the root only the upper sign ($-$) to satisfy the initial condition $%
\mathcal{E}\left( \tau \right) =\mathcal{E}_{0}$ when $\xi (\tau )=0$. In
the case $n=1$, Eq. (\ref{5}) yields known vacuum result. Further
investigation is devoted to the case of a medium with refraction index $n>1$%
. In this case the nature of the particle motion essentially depends on the
initial conditions and the value of the parameter $\xi \left( \tau \right) $
as far as the expression under the root in Eq. (\ref{5}) may become
negative. To solve this problem one needs to pass the complex plane were Eq.
(\ref{5}) has the form of known inverse Jukowski function.

If $\xi _{max}<\xi _{cr}$ ($\xi _{max}$ is the maximum value of the
parameter $\xi (\tau )$) the expression under the root in Eq. (\ref{5}) is
always positive and in front of the root one has to take the upper sign ($-$%
) according to the initial condition. Then $\mathcal{E}=\mathcal{E}_{0}$
after the interaction ($\xi (\tau )\rightarrow 0$) and the particle energy
remains unchanged.

If $\xi _{max}>\xi _{cr}$ the particle is unable to penetrate into the wave,
i.e., into the region $\xi >\xi _{cr}$ since at $\xi >\xi _{cr}$ the root in
Eq. (\ref{5}) becomes a complex one. This complexity now is bypassed via
continuously passing from one Riemann sheet to another, which corresponds to
changing the inverse Jukowski function from \textquotedblleft $-$%
\textquotedblright\ to \textquotedblleft $+$\textquotedblright\ before the
root. Hence, the upper sign ($-$) in this case stands up to the value of the
wave intensity $\xi (\tau )<\xi _{cr}$, then at $\xi \left( \tau \right)
=\xi _{cr}$ the root changes its sign from \textquotedblleft $-$%
\textquotedblright\ to \textquotedblleft $+$\textquotedblright , providing
continuous value for the particle energy in the field. The intensity value $%
\xi \left( \tau \right) =\xi _{cr}$ of the wave is a turn point for the
particle motion, so that we call it the critical value.

Thus, when the maximum value of the wave intensity exceeds the critical
value a transverse plane EM wave in the medium becomes a potential barrier
and the \textquotedblleft reflection\textquotedblright\ of the particle from
the wave envelope ($\xi (\tau )$) takes place. If now $\xi (\tau
)\rightarrow 0$, we obtain after the \textquotedblleft
reflection\textquotedblright\ for the particle energy 
\begin{equation}
\mathcal{E}=\mathcal{E}_{0}\left[ 1+2\frac{1-n\frac{\mathrm{v}_{0}}{c}}{%
n^{2}-1}\right] .  \label{7}
\end{equation}

If the initial conditions are such that the wave pulse overtakes the
particle ( $\mathrm{v}_{0}<c/n$ ), then after the \textquotedblleft
reflection\textquotedblright\ $\mathcal{E}>\mathcal{E}_{0}$ and the particle
is accelerated. But if the particle overtakes the wave ( $\mathrm{v}_{0}>c/n$%
), then $\mathcal{E}<\mathcal{E}_{0}$ and particle deceleration takes place.

This threshold phenomenon of the particle \textquotedblleft
reflection\textquotedblright\ can be more clearly presented in the frame of
reference connected with the wave. In this frame the electric field of the
wave vanishes ($\mathbf{E}^{\prime }\equiv 0$) and there is only the static
magnetic field ($\left\vert \mathbf{H}^{\prime }\right\vert =\left\vert 
\mathbf{H}\right\vert \sqrt{n^{2}-1}/n$ ).

The phenomenon of charged particle \textquotedblleft
reflection\textquotedblright\ from a plane EM wave may also be used for the
monochromatization of particle beams. The fact that above the critical
intensity value the induced Cherenkov process occurs in only one direction
-- either emission or absorption -- and for the initial Cherenkov velocity $%
\mathrm{v}_{0x}=c/n$ the energy of the particle after the \textquotedblleft
reflection\textquotedblright\ does not change, in principle enables
conversion of the energetic or angular spreads of charged particle beams. In
this case the energy of the particle is given by Eq. (\ref{1}), which at the
actual values of the parameters for induced Cherenkov process the second
term under the root is much smaller than the third one, that is, $%
2p_{0}|\sin \vartheta |/mc\gg \xi _{max}$ and for the critical field in this
case (at the wave linear polarization - electric field strength along the
axis $OY$) we have 
\begin{equation}
\xi _{cr}(\vartheta )=\frac{c}{2\mathrm{v}_{0}}\frac{\mathcal{E}_{0}}{mc^{2}}%
\frac{\left( 1-n\frac{\mathrm{v}_{0}}{c}\cos \vartheta \right) ^{2}}{\left(
n^{2}-1\right) |\sin \vartheta |}\ ;\qquad \vartheta \neq 0  \label{9}
\end{equation}%
(in the case $\vartheta =0$, $\xi _{cr}$ is determined by Eq. (\ref{6})).

If the maximal value of the wave intensity $\xi _{max}>\xi _{cr}(\vartheta )$%
, then the particle energy after the \textquotedblleft
reflection\textquotedblright\ is 
\begin{equation}
\mathcal{E}\left( \vartheta \right) =\mathcal{E}_{0}\left[ 1+\frac{2\left(
1-n\frac{\mathrm{v}_{0}}{c}\cos \vartheta \right) }{n^{2}-1}\right] .
\label{10}
\end{equation}

Let the charged particle beam with an initial energetic ( $\Delta _{0}$) and
angular ($\delta _{0}$) spread interact with a plane transverse EM wave of
intensity $\xi _{max}>\xi _{cr}(\vartheta )$ in a gaseous medium. To keep
the mean energy $\overline{\mathcal{E}_{0}}$ of the beam unchanged after the
interaction (at the adiabatic turning on and turning off of the wave) the
axis of the beam with mean velocity $\overline{\mathrm{v}_{0}}$ must be
pointed at the Cherenkov angle ($\vartheta _{0}$) to the laser beam, i.e., $%
n(\overline{\mathrm{v}_{0}}/c)\cos \vartheta _{0}=1$. Under this condition
the particles with velocities $\mathrm{v}_{0}\cos \vartheta <c/n$ will
acquire an energy and the other particles for which the longitudinal
velocities exceed the phase velocity of the wave ($\mathrm{v}_{0}\cos
\vartheta >c/n$) will loss an energy according to Eq. (\ref{10}). As a
result the energies of the particles $\mathcal{E}\left( \vartheta \right) $
will approach close to the mean energy $\overline{\mathcal{E}_{0}}$ of the
beam ($\mathcal{E}\left( \vartheta \right) \rightarrow $ $\overline{\mathcal{%
E}_{0}}$ ) and the final energetic width of the beam will become less than
the initial one. As there is one free parameter (for a specified velocity $%
\overline{\mathrm{v}_{0}}$ the parameters $\vartheta _{0}$ and $n$ are
related by Cherenkov condition) it is possible to use it to control the
exchange in the energy of the particles after the \textquotedblleft
reflection\textquotedblright\ (\ref{10}) and to reach the minimal final
energy spread of the beam $\Delta \ll \Delta _{0}$ -- monochromatization.
Depending on the relation between the initial energetic and angular spreads
and mean energy of the beam, the opposite process may occur, namely angular
narrowing of the beam. Physically it is clear that with the
monochromatization the angular divergence of the beam will increase and the
opposite -- the angular narrowing of the beam -- leads to
demonochromatization (in accordance with Liouville's theorem). More detailed
consideration of this effect with the quantitative results can be found in
the papers \cite{Ch._mon.,conv.} (for suppression of a beam spread by vacuum
versions of \textquotedblleft reflection\textquotedblright\ phenomenon see
the papers \cite{Comp._mon.,und._mon.}).

To illustrate the typical picture of nonlinear interaction of a charged
particle with a strong EM wave in a medium we present the graphics of
numerical solutions for the laser pulse of finite duration, showing the
behavior of particle dynamics below and above critical intensity, with the
effect of acceleration. At first we will not take into account the
dependence of the slowly varying intensity envelope of a laser beam from the
transversal coordinates. Thus, a laser beam may be modeled as 
\begin{equation}
E_{x}=0,\quad E_{z}=0,\quad E_{y}=\frac{E_{0}}{\cosh \left( \frac{\tau }{%
\delta \tau }\right) }\cos \omega \tau ,  \label{EY1}
\end{equation}%
where $\delta \tau $ characterizes the pulse duration. 
%%%%%%%%%%%%%%%%%%%%%%%%%%%%%%%%%%%%%%%%%%%%%%%%%%%%%%%%%%%%%%%%
\begin{figure}[b]
\includegraphics[width=.47\textwidth]{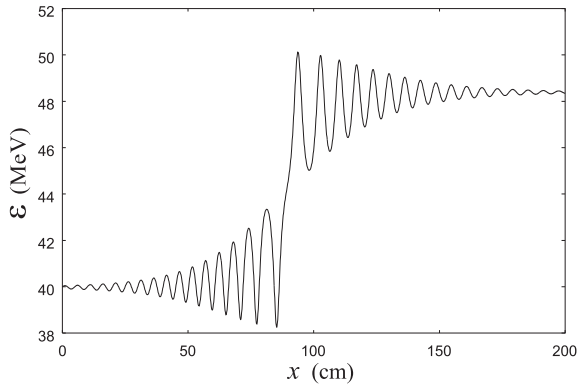}
\caption{\textquotedblleft Reflection\textquotedblright\ of the particle.
The energy versus the position $x$ is plotted when the wave intensity is
above the critical point.}
\label{eps2.1}
\end{figure}
%%%%%%%%%%%%%%%%%%%%%%%%%%%%%%%%%%%%%%%%%%%%%%%%%%%%%%%%%%%%%%%%
The particle initial energy is taken to be $\mathcal{E}_{0}=40$ $\mathrm{MeV}
$ and the initial velocity is directed at the angle $\vartheta =9\times
10^{-3}$ $\mathrm{rad}$ to the wave propagation direction ($p_{0z}=0$). The
refraction index of the gaseous medium for this calculation has been chosen
to be $n-1=10^{-4}$. Figure~1 illustrates the evolution of the particle
energy: the energy versus the position $x$ is plotted for a neodymium laser (%
$\hbar \omega \simeq 1.17$ $\mathrm{eV}$) with electric field strength $%
E_{0}=3\times 10^{8}$ $\mathrm{V/cm}$ and $\delta \tau =4T$ ($T$ is the wave
period). For these parameter values the wave intensity is above the critical
point and, as we see from this figure, the particle energy is abruptly
changed corresponding to the \textquotedblleft reflection\textquotedblright\
phenomenon. Figure 2a illustrates the evolution of the energies of particles
with different initial interaction angles. The initial energies for all
particles are $\mathcal{E}_{0}\simeq 40$ $\mathrm{MeV}$. Figure 2b
illustrates the role of initial conditions: the final energy versus the
interaction angle is plotted. As follows from Eq. (\ref{9}) the critical
intensity, as well as the final energy (\ref{10}), depend on the initial
interaction angle and as a consequence we have this picture. Note that the
acceleration rate neither depends on the field magnitude (only should be
above threshold field) nor on the interaction length.

%%%%%%%%%%%%%%%%%%%%%%%%%%%%%%%%%%%%%%%%%%%%%%%%%%%%%%%%%%%%%%%%
\begin{figure}[tbp]
\includegraphics[width=.48\textwidth]{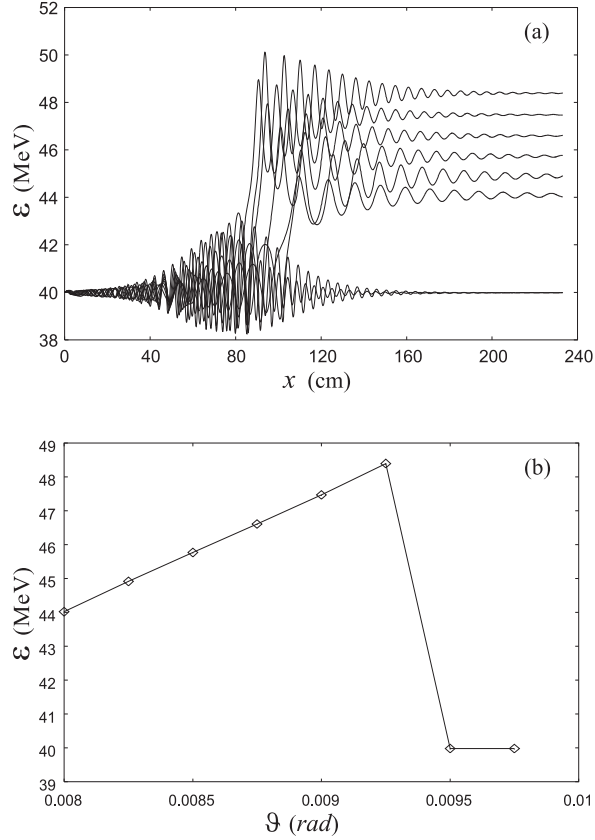}
\caption{\textquotedblleft Reflection\textquotedblright\ of the particles
with different initial interaction angles. Panel (a) displays the evolution
of the energies of particles. In (b) the final energy versus the interaction
angle is plotted.}
\label{eps2.2}
\end{figure}
%%%%%%%%%%%%%%%%%%%%%%%%%%%%%%%%%%%%%%%%%%%%%%%%%%%%%%%%%%%%%%%%
%%%%%%%%%%%%%%%%%%%%%%%%%%%%%%%%%%%%%%%%%%%%%%%%%%%%%%%%%%%%%%%%
\begin{figure}[tbp]
\includegraphics[width=.48\textwidth]{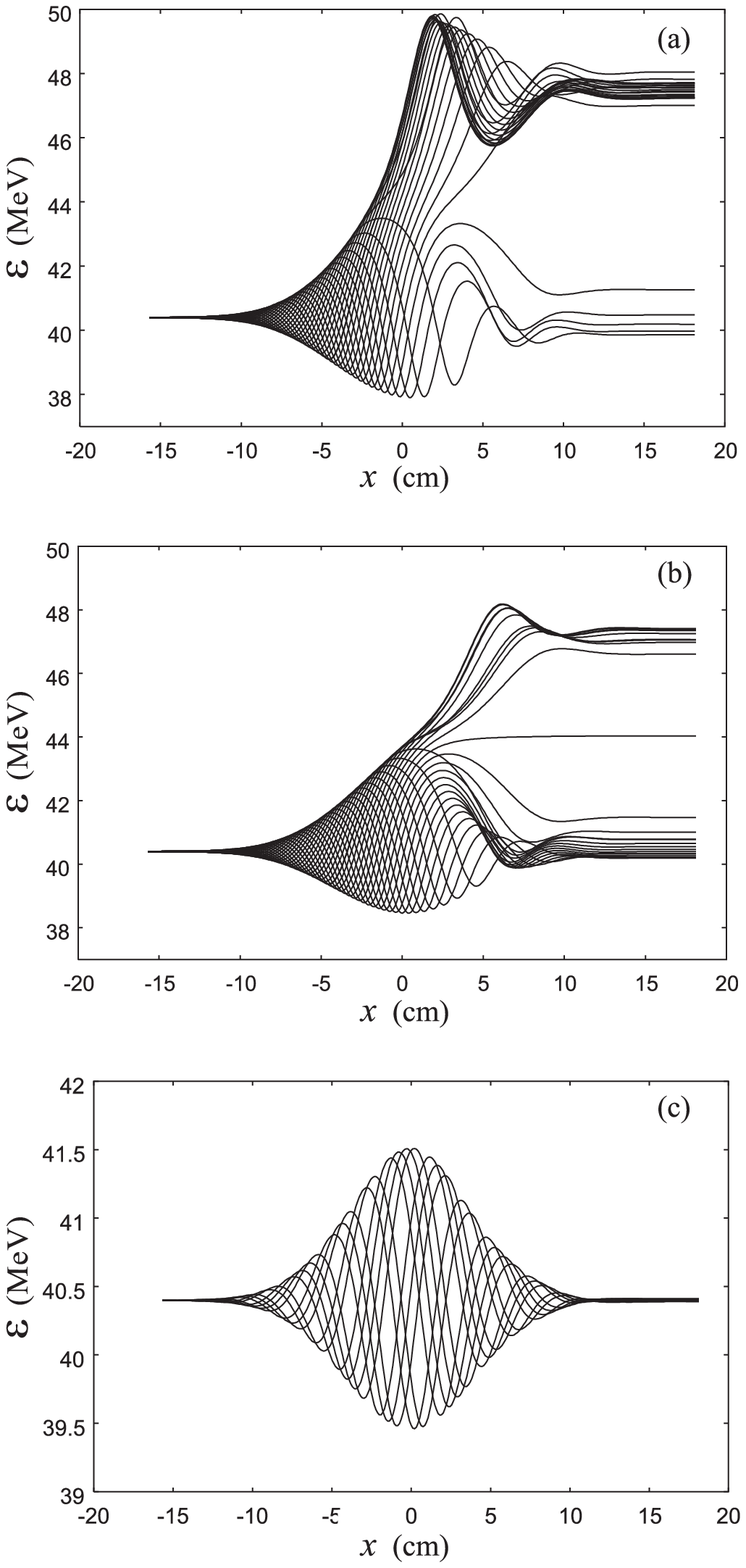}
\caption{The evolution of the energies of particles with various initial
phases are shown for the laser beam with transversal intensity profile for
the various entrance cordinates: (a) $z=0$, (b) $z=d/4$, and (c) $z=d/2$.}
\label{eps2.3}
\end{figure}
%%%%%%%%%%%%%%%%%%%%%%%%%%%%%%%%%%%%%%%%%%%%%%%%%%%%%%%%%%%%%%%%
To demonstrate the dependence of the considered process on transversal
profile of the laser intensity for actual beams in Fig. 3 the evolution of
the energies of particles with various initial phases (with initial energies 
$\mathcal{E}_{0}\simeq 40$ $\mathrm{MeV}$) is illustrated. The laser beam
transversal profile is modeled by the Gaussian function 
\begin{equation}
E_{y}=E_{0}\exp \left( -\frac{4}{d^{2}}\left( y^{2}+z^{2}\right) \right) 
\frac{\cos \omega \tau }{\cosh \left( \frac{\tau }{\delta \tau }\right) }
\label{EY2}
\end{equation}%
with $d=10^{3}\lambda ,$ $\delta \tau =50T$. As we see from this figure the
acceleration picture is essentially changed depending on the entrance
coordinates of the particles. This is the manifestation of the threshold
nature of the \textquotedblleft reflection\textquotedblright\ phenomenon.

If for the intensity exceeding the critical value a plane EM wave becomes a
potential barrier for the external particle (with respect to the wave), then
for the particle initially situated in the wave it may become a potential
well and particle capture by the wave will take place \cite{2006}.

\subsection{Laser Acceleration in Gaseous Media. Cherenkov Accelerator}

The phenomenon of charged particle \textquotedblleft
reflection\textquotedblright\ and capture by a transverse EM wave can be
used for particle acceleration in laser fields. As the application of large
intensities in this process is restricted because of the medium ionization
the acceleration owing to \textquotedblleft reflection\textquotedblright\ in
the medium with refraction index $n=\mathrm{const}$ -- single
\textquotedblleft reflection\textquotedblright -- is relatively small.
However, if the refraction index decreases along the wave propagation
direction in such a way that the condition of particle synchronous motion
with the wave $\mathrm{v}_{x}(x)=c/n(x)$ takes place continuously, the phase
velocity of the wave will increase all the time and the particle being in
front of the wave barrier (at $\xi >\xi _{cr}$) will continuously be
\textquotedblleft reflected\textquotedblright , i.e., continuously
accelerated. The law $n=n(x)$ must have an adiabatic character not to allow
the particle to leave the wave after the single \textquotedblleft
reflection\textquotedblright . Such variation law of the refraction index
can be realized in a gaseous medium adiabatically decreasing the pressure.

For particle acceleration one can also use the capture regime. In this case
in the medium with $n=\mathrm{const}$ the particle energy does not change on
average (particle makes stable oscillations around the equilibrium phases in
the wave moving with average velocity $<\mathrm{v}_{x}>=c/n$). However, if
one decreases the refraction index along the propagation direction of the
wave, so that the particle does not leave the equilibrium phases, then the
wave will continuously accelerate the particle. Then, to realize the capture
regime one needs $p_{0y}/mc\xi <2$. For not very strong fields this is
sufficiently strict confinement on the transverse momentum of the particle.
For particle acceleration by laser fields one can use the capture regime
corresponding to large transverse momenta of the particle $p_{0y}/mc\xi >2$.
So, we will consider the general case of particle capture with arbitrary
initial momentum $\mathbf{p}_{0}$ and laser acceleration in gaseous medium
with varying refraction index $n(x)$. The variation law of the medium
refraction index for acceleration is determined by equation: 
\begin{equation}
\frac{dn(x)}{dx}=-\frac{\omega }{c}\xi \frac{p_{ys}}{mc}\cos \phi _{s}\left( 
\frac{mc^{2}}{\mathcal{E}_{s}(x)}\right) ^{2}n^{2}(x)\left[ n^{2}(x)-1\right]
.  \label{62}
\end{equation}

It is seen from this equation that for the particle acceleration in the
capture regime via decreasing refraction index of the medium ($dn(x)/dx<0$)
one needs $p_{ys}\cos \phi _{s}>0$ (equilibrium transverse momentum of the
particle must be directed along the vector of the wave electric field). In
the opposite case the continuous deceleration of the particle will take
place accompanied by induced Cherenkov radiation (regime of continuous
amplification of the wave by the particle beam at $dn(x)/dx>0$).

The energy of equilibrium particle acquired on the distance $x$ is defined
by 
\begin{equation}
\mathcal{E}_{s}^{2}(x)=\frac{n^{2}(x)}{n^{2}(x)-1}\left(
m^{2}c^{4}+c^{2}p_{ys}^{2}\right) .  \label{63}
\end{equation}

Equation (\ref{63}) in the general case defines the particle acceleration in
the capture regime when the medium refraction index falls along the wave
propagation. It defines the longitudinal dimension of such \textquotedblleft
Cherenkov accelerator\textquotedblright\ as well. The transverse dimension
of the latter is defined by 
\begin{equation}
\mathcal{E}_{s}(y)=\mathcal{E}_{s}(0)+mc\omega \xi (y-y_{0})\cos \phi _{s}.
\label{65}
\end{equation}

Here $\mathcal{E}_{s}(0)$ and $y_{0}$ are the initial equilibrium values of
the energy and transverse coordinate of the particle ($y-y_{0}$ is the
transverse dimension of \textquotedblleft Cherenkov
accelerator\textquotedblright ). As is seen from Eq. (\ref{65}) the particle
acceleration takes place if $(y-y_{0})\cos \phi _{s}>0$, and in the opposite
case deceleration occurs ($\mathcal{E}_{s}(y)<\mathcal{E}_{s}(0)$) in
accordance with what was mentioned above. For relativistic particles, when $%
n(x)\sim 1$ and $n(x)-1\ll n(0)-1$, from Eq. (\ref{62}) we have 
\begin{equation}
n(x)-1\simeq \frac{m^{2}c^{4}+c^{2}p_{ys}^{2}}{4mc^{2}\xi \omega p_{ys}\cos
\phi _{s}}\frac{1}{x}.  \label{66}
\end{equation}

For energy of the equilibrium particle at large acceleration we have 
\begin{equation}
\mathcal{E}_{s}(x)\simeq \sqrt{2mc^{2}\xi \omega |p_{ys}\cos \phi _{s}|x}\
;\qquad \mathcal{E}_{s}(x)\gg \mathcal{E}_{s}(0).  \label{69}
\end{equation}

The estimations show that, for example, at electric field strengths of laser
radiation $E\sim 10^{8}$ $\mathrm{V/cm}$ an electron with initial energy $%
\mathcal{E}_{s}(0)\sim 5$ $\mathrm{MeV}$ acquires energy $\mathcal{E}%
_{s}(x)\sim 50$ $\mathrm{MeV}$ already at the distance $x\sim 1$ $\mathrm{cm}
$. The transverse dimension of acceleration $y-y_{0}\ $is of the order of a
few millimeters and the longitudinal dimension of the system is of the order
of the transverse one (a few times larger). At the distance $x\sim 1$ $%
\mathrm{m}$ the particle energy gain is of the order of $1$ $\mathrm{GeV}$ .
Note that because of multiple scattering on the atoms of the medium the
particles can leave the regime of stable motion as a result of change of $%
p_{ys}$. The analysis shows that the multiple scattering essentially falls
in the above-mentioned gaseous media for laser field strengths $E>10^{7}$ $%
\mathrm{V/cm}$.

To illustrate the particle acceleration in the capture regime we will
represent the results of numerical solution of relativistic equations of
motion in the field of an actual laser beam with the electric field strength 
\begin{equation}
\quad E_{y}=E_{0}\exp \left( -\frac{4}{d^{2}}\left( y^{2}+z^{2}\right)
\right) \frac{\cos \left( \frac{\omega }{c}\int n(x)dx-\omega t+\varphi
_{0}\right) }{\cosh \left( \frac{\frac{1}{c}\int n(x)dx-t+\varphi
_{0}/\omega }{\delta \tau }\right) },  \label{EY3}
\end{equation}%
\begin{equation*}
E_{x}=0,\quad E_{z}=0,
\end{equation*}%
where $\delta \tau $ characterizes the pulse duration and $\varphi _{0}$ is
the initial phase. Simulations have been made for neodymium laser ($\hbar
\omega \simeq 1.17$ $\mathrm{eV}$) with electric field strength $%
E_{0}=3\times 10^{8}$ $\mathrm{V/cm}$ and $\delta \tau =1000T,$ $d=5\times
10^{3}\lambda $. The variation law for the refraction index of the medium is
defined in self-consistent manner that may be approximated by the function 
\begin{equation}
n(x)=\frac{n_{0}+n_{f}}{2}+\frac{\left( n_{f}-n_{0}\right) }{2}\tanh \left(
\kappa x\right) ,  \label{NX4}
\end{equation}%
where $n_{0}$, $n_{f}$ are the initial and final values of the refraction
index and $\kappa $ characterizes the decreasing rate.

%%%%%%%%%%%%%%%%%%%%%%%%%%%%%%%%%%%%%%%%%%%%%%%%%%%%%%%%%%%%%%%%
\begin{figure}[tbp]
\includegraphics[width=.45\textwidth]{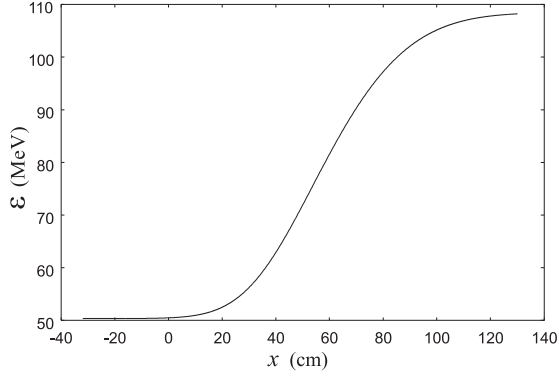}
\caption{The evolution of the particle energy in the capture regime with
variable refractive index.}
\label{eps2.4}
\end{figure}
%%%%%%%%%%%%%%%%%%%%%%%%%%%%%%%%%%%%%%%%%%%%%%%%%%%%%%%%%%%%%%%%

%%%%%%%%%%%%%%%%%%%%%%%%%%%%%%%%%%%%%%%%%%%%%%%%%%%%%%%%%%%%%%%%
\begin{figure}[tbp]
\includegraphics[width=.48\textwidth]{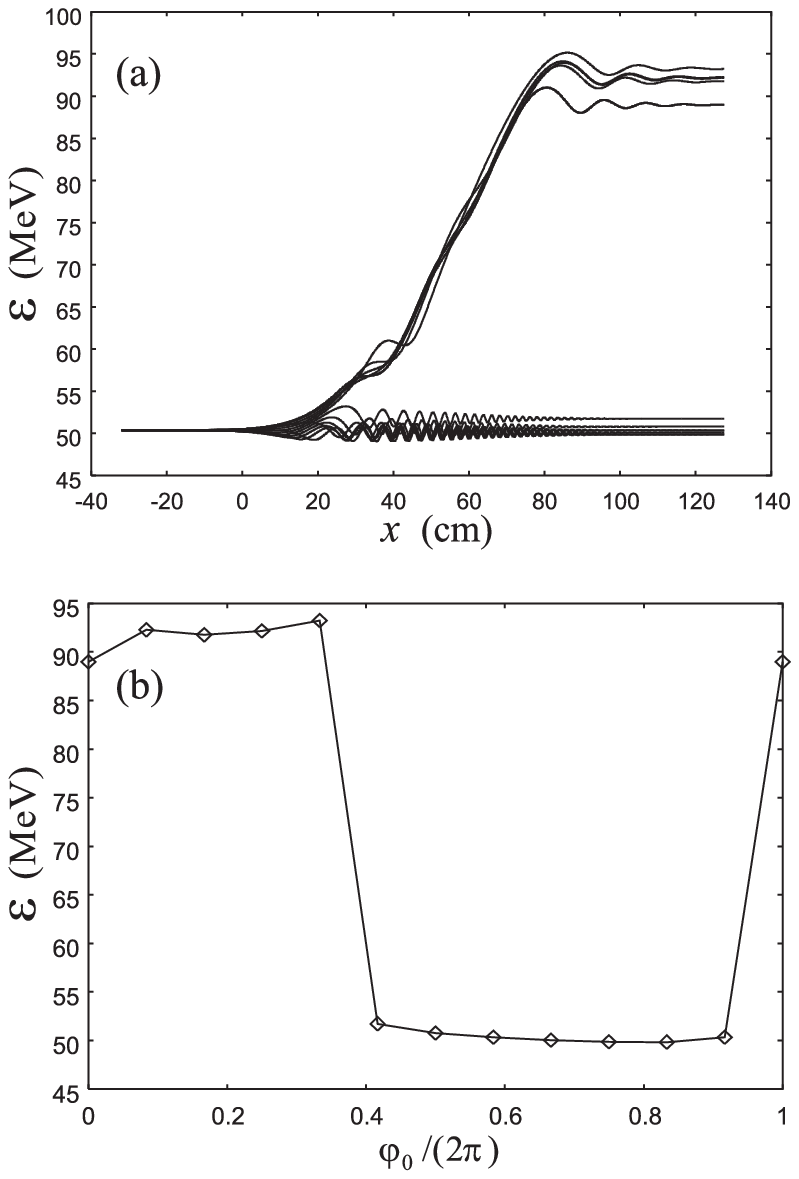}
\caption{Acceleration of the particles in the capture regime. Panel (a)
displays the evolution of the energies of particles with various initial
phases. The initial entrance coordinate is $z=0$. In (b) the final energy
versus the initial phase is plotted.}
\label{eps2.5}
\end{figure}
%%%%%%%%%%%%%%%%%%%%%%%%%%%%%%%%%%%%%%%%%%%%%%%%%%%%%%%%%%%%%%%%

%%%%%%%%%%%%%%%%%%%%%%%%%%%%%%%%%%%%%%%%%%%%%%%%%%%%%%%%%%%%%%%%
\begin{figure}[tbp]
\includegraphics[width=.48\textwidth]{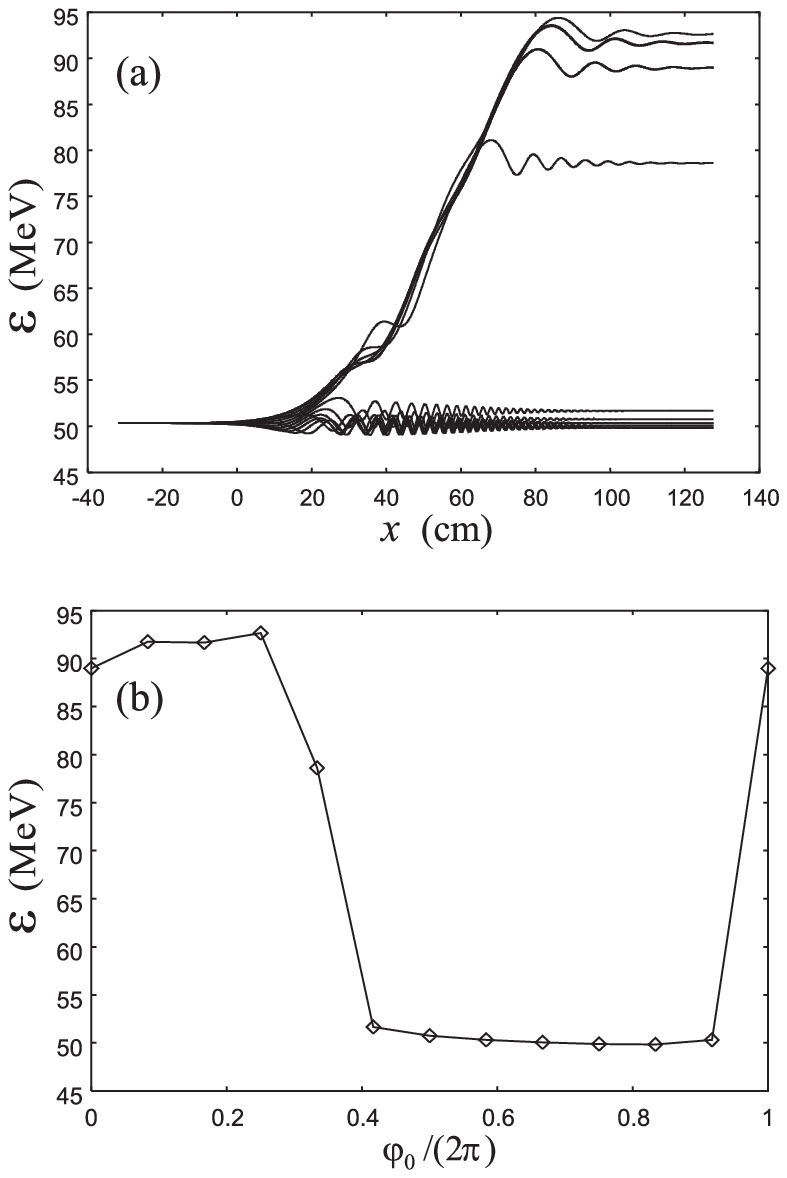}
\caption{Acceleration of the particles in the capture regime. Panel (a)
displays the evolution of the energies of particles with various initial
phases. The initial entrance coordinate is $z=0.25$ $\mathrm{mm}$. In (b)
the final energy versus the initial phase is plotted.}
\label{eps2.6}
\end{figure}
%%%%%%%%%%%%%%%%%%%%%%%%%%%%%%%%%%%%%%%%%%%%%%%%%%%%%%%%%%%%%%%%

Figure~4 illustrates the evolution of the particle energy in the capture
regime. The particle initial energy is taken to be $\mathcal{E}_{0}=50.5$ $%
\mathrm{MeV}$ and the initial velocity is directed at the angle $\vartheta
=9\times 10^{-3}$ $\mathrm{rad}$ to the wave propagation direction ($p_{0z}=0
$). The initial value of the refraction index has been chosen to be $%
n-1\simeq 10^{-4}$. As we see in the capture regime with variable refraction
index, one can achieve considerable acceleration.

To show the role of initial conditions in Fig. 5a the evolution of the
energies of particles with the same initial energies $\mathcal{E}_{0}=50.5$ $%
\mathrm{MeV}$ ($\vartheta =9\times 10^{-3}$ $\mathrm{rad}$) and various
initial phases is illustrated. The initial entrance coordinate is $z=0$.
Figure 5b displays the role of initial conditions: the final energy versus
the initial phase is plotted. In Fig.~6 the parameters are the same as in
Fig. 5a except the initial entrance coordinate, which is taken to be $z=0.25$
$\mathrm{mm}$. As we see from these figures the captured particles are
accelerated, while the particles situated in the unstable phases (or if the
conditions for capture are not fulfilled) after the interaction remain with
the initial energy.

\subsection{Cyclotron Resonance in a Medium. Nonlinear Threshold Phenomenon
of \textquotedblleft Electron Hysteresis\textquotedblright\ }

Consider now the dynamics of a charged particle in the field of a strong EM
wave in a medium at the presence of a static uniform magnetic field $\mathbf{%
H}_{0}$ along the wave propagation direction -- nonlinear cyclotron
resonance in a medium. In this case the problem can be solved analytically
only for the circular polarization of monochromatic wave and if the initial
velocity of the particle is directed along the axis of the wave propagation.

As far as the equation for the particle longitudinal momentum is not changed
in the presence of a uniform magnetic field with respect to equation in the
field of a plane EM wave in a medium, and the equation for the particle
energy change in the field remains unchanged, then one can represent the
particle longitudinal velocity $\mathrm{v}_{x}$ and energy 
\begin{equation}
\mathcal{E=}\frac{\mathcal{E}_{0}}{n^{2}-1}\left\{ n^{2}\left( 1-\frac{%
\mathrm{v}_{0}}{cn}\right) -\left( 1-n\frac{\mathrm{v}_{0}}{c}\right) \left[
1\mp \frac{\mathbf{p}_{\perp }^{2}(\tau )}{\left( mc\zeta \right) ^{2}}%
\right] ^{1/2}\right\}   \label{4.54}
\end{equation}%
via the transversal momentum $\mathbf{p}_{\perp }(\tau )=\left\{
0,p_{y}(\tau ),p_{z}(\tau )\right\} $ in the field. Here the parameter $%
\zeta $ is 
\begin{equation}
\zeta \equiv \frac{\mathcal{E}_{0}}{mc^{2}}\frac{\left\vert 1-n\frac{\mathrm{%
v}_{0}}{c}\right\vert }{\sqrt{\left\vert n^{2}-1\right\vert }}.  \label{4.56}
\end{equation}%
Note that $\zeta $ is the critical value of the wave intensity (\ref{6}) (at 
$n>1$) for the particle \textquotedblleft reflection\textquotedblright\
phenomenon in the absence of a static magnetic field ($H_{0}=0$). The sign
\textquotedblleft $-$\textquotedblright under the root in (\ref{4.54})
corresponds to the case of the interaction in dielectriclike media with $n>1$
and the sign \textquotedblleft $+$\textquotedblright , plasmalike media with 
$n<1$. Note that in contrast to the case $H_{0}=0$ (induced Cherenkov
process) in Eq. (\ref{4.54}) before the root, only the sign
\textquotedblleft $-$\textquotedblright\ is taken (in accordance with the
initial conditions $\mathrm{v}_{x}=\mathrm{v}_{0}$ and $\mathcal{E=E}_{0}$
of the free particle) since, as will be shown below, in this case the
expression under the root is always positive and consequently the root
cannot change its sign. 

Now the considered problem reduces to definition of the particle transversal
momentum $\mathbf{p}_{\perp }(\tau )$ which in the wave coordinate $\tau =t-$
$nx/c$ obey the following equation: 
\begin{equation}
\frac{dZ\left( \tau \right) }{d\tau }=\frac{eE(\tau )}{mc}-i\frac{\Omega }{%
\left( 1-n\frac{\mathrm{v}_{0}}{c}\right) \left[ 1\mp \frac{\left\vert
Z\left( \tau \right) \right\vert ^{2}}{\zeta ^{2}}\right] ^{1/2}}Z\left(
\tau \right) ,  \label{4.57}
\end{equation}
where the complex quantity $Z\left( \tau \right) $ related to the
dimensionless parameter of the particle transversal momentum%
\begin{equation}
Z\left( \tau \right) =\frac{p_{y}(\tau )+ip_{z}(\tau )}{mc}  \label{4.57a}
\end{equation}%
$E(\tau )=E_{y}(\tau )+iE_{z}(\tau )$ and $\Omega =ecH_{0}/\mathcal{E}_{0}$
is the Larmor frequency for the initial velocity of the particle.

For an arbitrary plane EM wave Eq. (\ref{4.57}) is a nonlinear equation the
exact solution of which cannot be found. However, for the monochromatic wave
of circular polarization when $E(\tau )=E_{0}\exp (-i\omega \tau )$ one can
find the exact solution\ of Eq. (\ref{4.57}). The latter is sought in the
form $Z(\tau )=Z_{0}\exp (-i\omega \tau )$ and for the transversal momentum
of the particle we obtain the following algebraic equation: 
\begin{equation}
\left( 1-\frac{\Omega }{\omega \left( 1-n\frac{\mathrm{v}_{0}}{c}\right) 
\sqrt{1\mp \beta ^{2}}}\right) \beta =X,  \label{4.60}
\end{equation}%
where the quantities $E_{0}$, $Z_{0}$ are expressed in the scale of the
parameter $\zeta $: 
\begin{equation}
\frac{Z_{0}}{\zeta }\equiv i\beta ;\quad \frac{eE_{0}}{mc\omega \zeta }=%
\frac{\xi _{0}}{\zeta }\equiv X.  \label{4.61}
\end{equation}

We will not represent here the exact solution of Eq. (\ref{4.60}) for $\beta 
$. An interesting nonlinear phenomenon exists in this process \cite{Hyst}
which can be found out through the graphical solution of Eq. (\ref{4.60}).
Thus, depending on the ratio of the Larmor and wave frequencies as well as
on the initial velocity of the particle (in the case of dielectriclike
medium where $\mathrm{v}_{0}\lessgtr c/n$) the solution of Eq. (\ref{4.60})
is a single-valued or multivalent that essentially changes the interaction
behavior of the particle with a strong EM wave at the nonlinear cyclotron
resonance in a medium. Hence, we will consider separately the cases $\Omega
\geqslant \omega ^{\prime }$ and $\Omega <\omega ^{\prime }$ at$\ \mathrm{v}%
_{0}<c/n$ where 
\begin{equation}
\omega ^{\prime }=\omega \left( 1-n\frac{\mathrm{v}_{0}}{c}\right)
\label{4.62}
\end{equation}%
is the Doppler-shifted frequency of the wave for the initial velocity of the
particle. If $\mathrm{v}_{0}>c/n$ the effects considered here will take
place with the opposite circular polarization of the wave ($\omega
\rightarrow -\omega $) or in the opposite direction of the uniform magnetic
field ($\mathbf{H}_{0}\rightarrow -\mathbf{H}_{0}$).

%%%%%%%%%%%%%%%%%%%%%%%%%%%%%%%%%%%%%%%%%%%%%%%%%%%%%%%%%%%%%%%%
\begin{figure}[tbp]
\includegraphics[width=.4\textwidth]{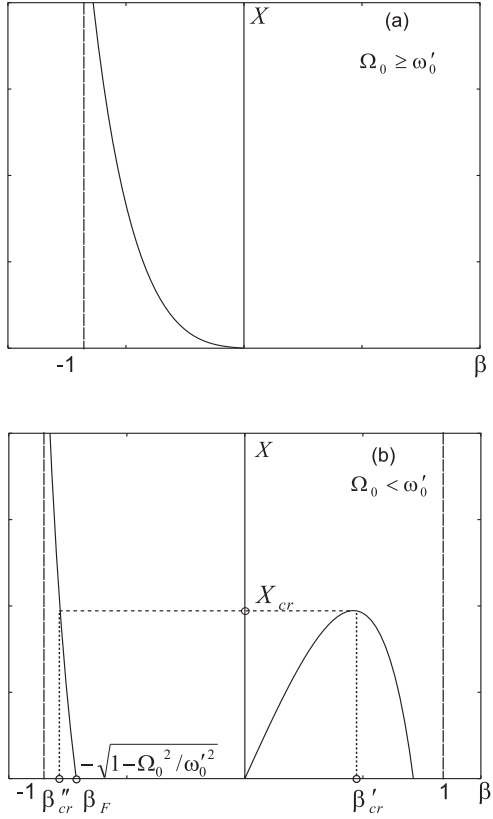}
\caption{Dependence of normalized transversal momentum $\protect\beta$ on
the normalized EM wave amplitude $X$ at $n_{0}>1$.}
\label{eps4.2}
\end{figure}
%%%%%%%%%%%%%%%%%%%%%%%%%%%%%%%%%%%%%%%%%%%%%%%%%%%%%%%%%%%%%%%%

Consider first the case of a medium with refractive index $n>1$ (sign
\textquotedblleft $-$\textquotedblright under the root) in Eq. (\ref{4.60}).
We will turn on the EM wave adiabatically and draw the graphic of dependence
of the particle transversal momentum on the wave intensity $\beta (X)$. For
the case $\Omega \geqslant \omega ^{\prime }$ the latter is illustrated in
Fig. 7a. As is seen from this graphic with the increase of the wave
intensity the transversal momentum of the particle increases in the field
(consequently the energy as well) and vice versa: with the decrease of the
wave intensity it decreases in the field and after the passing of the wave ($%
X=0$) the transversal momentum becomes zero ($\beta =0$), i.e., the particle
momentum-energy remain unchanged: $p=p_{0}$ and $\mathcal{E=E}_{0}$.

With the increase of the transversal momentum the longitudinal velocity of
the particle increases as well, but in contrast to the case $\mathbf{H}%
_{0}=0 $ it always remains smaller than the wave phase velocity if initially
the wave overtakes the particle ($\mathrm{v}_{0}<c/n$) and larger if the
particle overtakes the wave ($\mathrm{v}_{0}>c/n$). For this reason the
particle \textquotedblleft reflection\textquotedblright\ phenomenon vanishes
in the presence of a uniform magnetic field. Indeed, as is seen from Eq. (%
\ref{4.60}) for an arbitrary finite value of $X$ we have $\beta <1$ and
longitudinal velocity of the particle in the field $\mathrm{v}_{x}<c/n$ if $%
\mathrm{v}_{0}<c/n$ and $\mathrm{v}_{x}>c/n$ if $\mathrm{v}_{0}>c/n$. The
value $\beta =1$ may be reached only at $X=\infty $ when the root in Eq. (%
\ref{4.60}) becomes zero and $\mathrm{v}_{x}=c/n$. So, the expression under
the root in Eq. (\ref{4.60}) cannot become zero for finite intensities of
the EM wave and, consequently, the root cannot change its sign.

Consider now the case $\Omega <\omega ^{\prime }$. The graphic of dependence
of the particle transversal momentum on the wave intensity $\beta (X)$ in
this case is illustrated in Fig. 7b. As is seen from this graphic $\beta (X)$
is already a multivalent function: for wave intensities smaller than the
value corresponding to the maximum point of the curve $\beta (X)$ three
values of the particle transversal momentum exist for each value of the wave
intensity. At the maximum point, which will be called a critical one, the
wave intensity has the value 
\begin{equation}
X_{cr}=\left[ 1-\left( \frac{\Omega }{\omega ^{\prime }}\right) ^{2/3}\right]
^{3/2}.  \label{4.63}
\end{equation}%
There are two values $\beta _{cr}^{\prime }$ and $\beta _{cr}^{\prime \prime
}$ which correspond to critical intensity (\ref{4.63}). The first one, $%
\beta _{cr}^{\prime }$, is the value of the parameter $\beta $ corresponding
to particle transversal momentum at the maximum point of the curve $\beta (X)
$. From the extremum condition of Eq. (\ref{4.60}) for $\beta _{cr}^{\prime }
$ we have 
\begin{equation}
\beta _{cr}^{\prime }=\left[ 1-\left( \frac{\Omega }{\omega ^{\prime }}%
\right) ^{2/3}\right] ^{1/2}.  \label{4.64}
\end{equation}%
The second critical value for the parameter $\beta $ corresponding to
critical intensity $X_{cr}$ is situated on the left-hand side branch of the
curve $\beta (X)$. To determine its value one needs the analytic solution $%
\beta =\beta (X)$ of Eq. (\ref{4.60}), but there is no necessity here to
present the bulk expression for $\beta _{cr}^{\prime \prime }$.

We shall decide on that branch of the curve $\beta (X)$ which corresponds to
real motion of the particle. Up to the critical point the particle
transversal momentum can be changed on that branch which corresponds to
initial condition $\beta =0$ at $X=0$. On this branch the particle momentum
increases with the increase of the wave intensity and vice versa. It is
evident that with further increase of the field the particle cannot be
situated on the right-hand side from the critical point. Hence, it should
pass to the left-hand side branch of the curve $\beta (X)$. Indeed, it is
easy to see that the critical point is an unstable state for the particle,
while all states on the left-hand side branch of the curve $\beta (X)$ are
stable and at the critical point the particle changes instantaneously its
transversal momentum and passes by jumping to that branch. The further
variation of the particle transversal momentum occurs already on this
branch. Note that the instantaneity here is related to the fact that the
solution of Eq. (\ref{4.57}) has been found for the monochromatic wave. It
is clear that the momentum change actually occurs during finite time. This
jump variation of the particle momentum (energy) is due to the induced
resonant absorption of energy from the wave at the critical point because of
which the particle state at this point becomes unstable and it leaves the
resonance point for a stable state that corresponds to the transversal
momentum $\beta _{cr}^{\prime \prime }$ on the left-hand side branch of the
curve $\beta (X)$. Indeed, if one draws a graphic of the dependence of the
particle transversal momentum on the ratio of the Larmor and wave
frequencies $\Omega /\omega ^{\prime }$ for a certain intensity of the wave
(Fig.~8), then it will be seen from the graphic $\beta (\Omega /\omega
^{\prime })$ that the cyclotron resonance in the strong EM wave field takes
place at the critical point with the satisfaction of the condition $\Omega
<\omega ^{\prime }$. The latter means that to reach the cyclotron resonance
in a medium, in contrast to vacuum autoresonance it is necessary to be
initially under the resonance condition, since due to the effect of the
strong wave field in a medium with refractive index $n>1$ the Larmor
frequency increases in the field and then reaches the resonance value. In
vacuum the cyclotron resonance proceeds at $\Omega =\omega ^{\prime }$ which
survives infinitely, because of which the energy of the particle turns to
infinity. Thus, from Eq. (\ref{4.60}) in this case ($n=1$) for the particle
transversal momentum we have 
\begin{figure}[tbp]
\includegraphics[width=.48\textwidth]{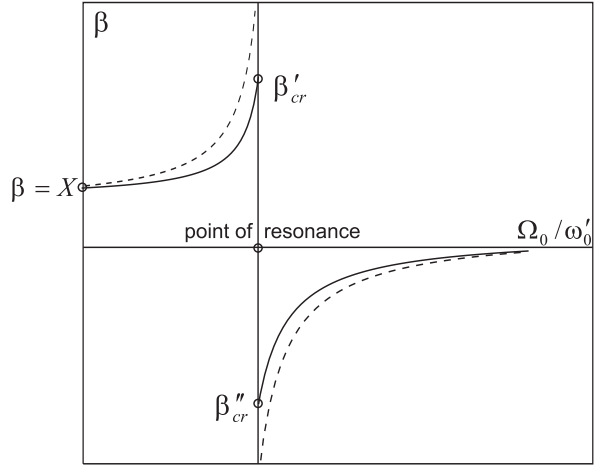}
\caption{Dependence of normalized transversal momentum on parameter $\Omega
_{0}/\protect\omega _{0}^{\prime }<1$ at $n_{0}>1.$}
\label{eps4.3}
\end{figure}
%%%%%%%%%%%%%%%%%%%%%%%%%%%%%%%%%%%%%%%%%%%%%%%%%%%%%%%%%%%%%%%%
%
\begin{equation}
\beta =\frac{X}{1-\frac{\Omega }{\omega ^{\prime }}},  \label{4.65}
\end{equation}%
which diverges (consequently the energy as well) at $\Omega =\omega ^{\prime
}$. As is seen from Fig.~8 this divergence vanishes in a medium.

With the further increase of the field ($X>X_{cr}$) the transversal momentum
of the particle will continuously increase on the left-hand side branch of
the curve $\beta (X)$ and tend to value $-1$ at $X\rightarrow \infty $. With
the decrease of the field the transversal momentum decreases on this branch
and at $X=X_{cr}$ already has only the value $\beta _{cr}^{\prime \prime }$
since the value $\beta _{cr}^{\prime }$ corresponds to the unstable state at
the resonance point and now there is no reason for inverse transition from
the stable state to the unstable one. With the further decrease of the field
the transversal momentum decreases, but as is seen from Fig.~7 after the
interaction ($X=0$) the particle does not return to the initial state ($%
\beta =0$ at $X=0$) and remains with the final transversal momentum 
\begin{equation}
\beta _{F}=-\left[ 1-\left( \frac{\Omega }{\omega ^{\prime }}\right) ^{2}%
\right] ^{1/2}.  \label{4.66}
\end{equation}%
This is a nonlinear phenomenon of charged particle hysteresis \cite{Hyst} in
the cyclotron resonance with a strong EM wave in a medium at intensities
exceeding the threshold value (\ref{4.63}).

The energy acquired by the particle due to hysteresis is given by 
\begin{equation}
\mathcal{E=E}_{0}\left[ 1+\frac{\left( 1-n\frac{\mathrm{v}_{0}}{c}\right)
\left( 1-\frac{\Omega }{\omega ^{\prime }}\right) }{n^{2}-1}\right] .
\label{4.68}
\end{equation}%
If the wave intensity is smaller than the critical value (\ref{4.63}) the
energy of the particle oscillates in the field and after the interaction
remains unchanged (such solution only has been obtained in the paper \cite%
{Buchb}).

Equation (\ref{4.68}) determines the particle acceleration due to a strong
transversal EM wave at the cyclotron resonance with the powerful static
magnetic field in a gaseous medium ($n-1\ll 1$). Because of the latter one
can achieve the cyclotron resonance using optical (laser) radiation in a
medium with the refractive index $n>1$, since the Doppler shift for a wave
frequency $1-n\mathrm{v}_{0}/c$ (see Eq. (\ref{4.62})) in this case may be
arbitrarily small in contrast to vacuum, where the\ cyclotron resonance for
the existing powerful static magnetic fields is possible only in the
radio-frequency domain. On the other hand, the application of powerful laser
radiation for large acceleration of the particles in gaseous media is
confined by the ionization threshold of the medium.

Consider now the case of a plasmous medium ($n<1$). In Eq. (\ref{4.60}) this
case should take the sign \textquotedblleft $+$\textquotedblright under the
root at which the confinement for the particle transversal momentum,
existing in a dielectriclike medium, vanishes. In addition, the
above-considered behavior of the\ cyclotron resonance in a plasmous medium
takes place with the inverse relation between the initial Larmor and wave
frequencies $\Omega /\omega ^{\prime }$. Thus, at $\Omega \leqslant \omega
^{\prime }$ with the increase of the wave intensity the transversal momentum
of the particle increases in the field and vice versa: with the decrease of
the wave intensity it decreases in the field and after the passing of the
wave ($X=0$) the transversal momentum becomes zero ($\beta =0$), i.e., the
particle momentum-energy remain unchanged: $p=p_{0}$ and $\mathcal{E=E}_{0}$%
. The nonlinear phenomenon of particle hysteresis in a plasmous medium takes
place at $\Omega >\omega ^{\prime }$, since in a medium with refractive
index $n<1$ the Larmor frequency decreases in the field and then becomes
equal to the resonance value. 
\begin{figure}[tbp]
\includegraphics[width=.4\textwidth]{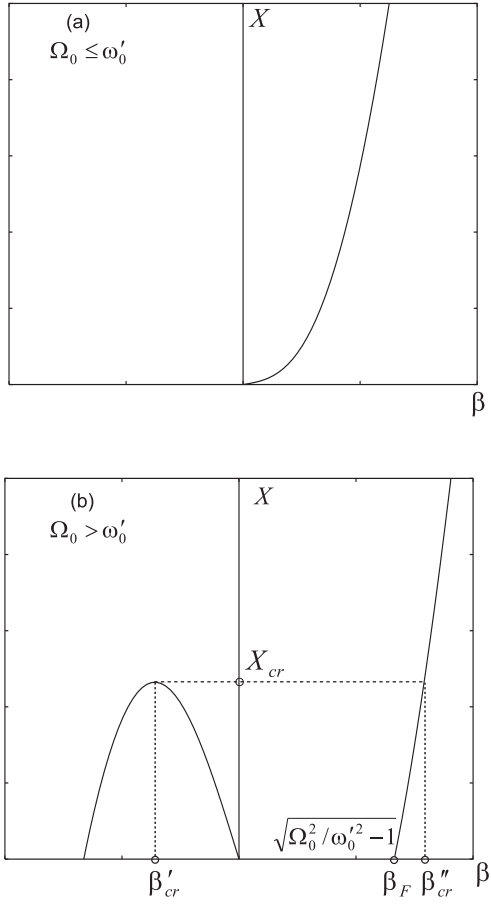}
\caption{Dependence of normalized transversal momentum $\protect\beta $ on
the normalized EM wave amplitude $X$ at $n_{0}<1$.}
\label{eps4.4}
\end{figure}
%%%%%%%%%%%%%%%%%%%%%%%%%%%%%%%%%%%%%%%%%%%%%%%%%%%%%%%%%%%%%%%%
The graphic of dependence of the particle transversal momentum on the wave
intensity $\beta (X)$ in this case is illustrated in Fig.~9. As is seen from
this graphic, in contrast to the case of dielectriclike media the parameter $%
\beta $ in the plasmas increases with no limit at the increase of the field.
The latter allows the large acceleration of the particles achieved by the
current superstrong laser fields of ultrarelativistic intensities ($\xi \gg 1
$) due to this phenomenon of hysteresis in the plasmas. The final
transversal momentum of the particle as a result of the hysteresis in this
case is 
\begin{equation}
\beta _{F}=\left[ \left( \frac{\Omega }{\omega ^{\prime }}\right) ^{2}-1%
\right] ^{1/2},  \label{4.69}
\end{equation}%
the final energy of which will be determined by the same equation (\ref{4.68}%
) since both the numerator and denominator of the fraction in the expression
analogous to Eq. (\ref{4.68}) for the particle energy in a plasma change
sign.

Note an interesting effect at the cyclotron resonance in a medium as well.
At $\Omega =\omega ^{\prime }$ no matter how weak the EM wave field is -- $%
\xi _{0}\ll $ $\zeta $ (that is, $\xi _{0}\ll 1$ even for $\zeta \sim 1$) --
from Eq. (\ref{4.60}) it follows that 
\begin{equation}
\left\vert \beta \right\vert \simeq \left( \frac{2\xi _{0}}{\zeta }\right)
^{1/3},  \label{4.70}
\end{equation}%
that is, an essential nonlinearity ($\sim \xi _{0}^{1/3}\gg \xi _{0}$)
arises in a case where one would expect a linear dependence on the field
according to linear theory. It is the consequence of nonlinear cyclotron
resonance the width of which is large enough in this case: 
\begin{equation}
\triangle \omega \simeq 2^{-1/3}\cdot \left( \frac{\xi _{0}}{\zeta }\right)
^{2/3}\omega ^{\prime }.  \label{4.71}
\end{equation}

\subsection{Coherent Radiation of Charged Particles in Capture Regime.
Cherenkov Amplifier}

The particle capture phenomenon may in principle serve as a FEL mechanism
(Cherenkov amplifier). For the latter one needs to solve the self-consistent
problem on the basis of the set of Maxwell--Vlasov equations. This problem
can be solved analytically if the particle initial velocity in the capture
regime is directed along the wave propagation and has a value close to the
Cherenkov one: $\mathrm{v}_{0}=\mathrm{v}_{0x}=\left( c/n_{0}\right) \left(
1+\mu \right) $; parameter $\mu \ll 1$.

The amplitude of the initial wave field should be a slowly varying function
of the space-time coordinates ($x$, $t$) with respect to the phase. The
problem is sensitive to the wave polarization, therefore it will be
investigated for both circular and linear polarizations. Consider at first
the wave Cherenkov amplification in the capture regime for the circular
polarization of stimulated wave with the boundary conditions $%
E_{y}(0,t)=E_{0}\cos \omega _{0}t$, $E_{z}(0,t)=-E_{0}\sin \omega _{0}t.$%
Related to particle we assume that it crosses the boundary of the medium $%
x=0 $ at the moment $t=t_{0}$. To define the electric current of the
particle stream we assume that the space is continuously filled with the
charged particles. Then at the moment $t_{0}$ in the point $x$ will be
situated only the particles for which $t_{0}=t-$ $n_{0}x/c$ (with accuracy
of a small parameter $\mu \ll 1$). Then for the self-consistent field we
obtain the equation \cite{capture}

Equation for self-consistent field has a simpler form over wave coordinates $%
\tau =t-n_{0}x/c$, $\eta =x$. For the field amplitude in these coordinates $%
E(t,x)=f(\tau ,\eta )$ we have 
\begin{equation}
\frac{\partial }{\partial \eta }f(\tau ,\eta )=\frac{2\pi e\rho _{0}}{%
n_{0}\left( n_{0}^{2}-1\right) ^{1/2}}\mu \sin \left[ \frac{e\left(
n_{0}^{2}-1\right) }{mc^{2}}\int_{0}^{\eta }f(\tau ,\eta ^{\prime })d\eta
^{\prime }\right] .  \label{37}
\end{equation}%
where $\rho _{0}$ is the mean density of the particles in the initial
stream, which will be assumed constant (since $\mu \ll 1$ the variation $%
\rho _{0}$ is small and can be neglected).The simple analytic solution can
be received at the incident monochromatic wave: $f(\tau ,0)=$ $E_{0}$. In
this case $f(\tau ,\eta )$ does not depend on $\tau $, i.e., $f(\tau ,\eta )=
$ $f(\eta )$, and for the quantity 
\begin{equation}
\varphi =\frac{e\left( n_{0}^{2}-1\right) }{mc^{2}}\int_{0}^{\eta }f(\eta
^{\prime })d\eta ^{\prime }  \label{38}
\end{equation}%
we have the nonlinear equation of anharmonic oscillator 
\begin{equation}
\varphi ^{\prime \prime }=\frac{2\pi e^{2}\rho _{0}\left( n_{0}^{2}-1\right)
^{1/2}}{mc^{2}n_{0}}\mu \sin \varphi ,  \label{39}
\end{equation}%
the general solution of which is the incomplete elliptic integral of the
first kind 
\begin{equation*}
\frac{1}{2}\left( n_{0}^{2}-1\right) \frac{eE_{0}x}{mc^{2}}%
=\int_{0}^{\varphi /2}\frac{dz}{\sqrt{1+\zeta ^{2}\sin ^{2}z}}\ ,
\end{equation*}%
\begin{equation}
\zeta ^{2}=\frac{8\pi \mu }{n_{0}\left( n_{0}^{2}-1\right) ^{3/2}}\frac{%
mc^{2}\rho _{0}}{E_{0}^{2}}.  \label{40}
\end{equation}

In the linear case when $\varphi \ll 1$ from Eq. (\ref{40}) we have \cite%
{capture} 
\begin{equation}
E(x)=E_{0}\left[ 
\begin{array}{l}
\cosh \left( \frac{x}{l_{c}}\right) ,\quad \mu >0, \\ 
\cos \left( \frac{x}{l_{c}}\right) ,\quad \mu <0.%
\end{array}%
\right.  \label{41}
\end{equation}%
Hence, for $\mu >0$, which corresponds to particles' initial velocity $%
\mathrm{v}_{0}>c/n_{0}$, exponential amplification of the incident wave
occurs. For $\mu <0$, that is, $\mathrm{v}_{0}<c/n_{0}$, the amplification
vanishes on average. The quantity in Eq. (\ref{41}) 
\begin{equation}
l_{c}=\left( \frac{mc^{2}n_{0}}{2\pi e^{2}\mu \rho _{0}\left(
n_{0}^{2}-1\right) ^{1/2}}\right) ^{1/2}  \label{42}
\end{equation}%
is the coherent length of amplification. Equation (\ref{37}) is an analogue
of the equation of the quantum amplifier. The role of inverse population in
atomic systems here performs detuning of the Cherenkov resonance $\mathrm{v}%
_{0}-c/n_{0}$ (parameter $\mu $).

Consider now the case of linear polarization of incident wave $%
E_{y}=E(x,t)\cos \left( \omega _{0}n_{0}x/c-\omega _{0}t\right) $. By
analogy with the previous case for the slowly varying amplitude of the
self-consistent field we have the equation 
\begin{equation*}
2is\omega _{0}\left( \frac{n_{0}}{c}\frac{\partial E_{s}}{\partial x}+\frac{%
n_{s}^{2}}{c^{2}}\frac{\partial E_{s}}{\partial t}\right) +\frac{s^{2}\omega
_{0}^{2}}{c^{2}}\left( n_{s}^{2}-n_{0}^{2}\right) E_{s}
\end{equation*}%
\begin{equation}
=i^{s}\frac{4\pi e\rho _{0}s\omega _{0}}{c\left( n_{0}^{2}-1\right) ^{1/2}}%
\mu J_{s}(\alpha ),  \label{49}
\end{equation}%
according to which all harmonics are radiated in contrast to circular
polarization of the wave.

Consider Eq. (\ref{49}) with regard to the presence and absence of
synchronism. In the last case, when $n_{s}\neq $ $n_{0}$ taking into account
the slow variation of the field amplitude from Eq. (\ref{49}) we obtain \cite%
{capture} 
\begin{equation}
E_{s}=i^{s}\mu \frac{4\pi ec\rho _{0}}{\left( n_{0}^{2}-1\right) ^{1/2}}%
\frac{1}{s\omega _{0}}\frac{1}{n_{s}^{2}-n_{0}^{2}}J_{s}(\alpha ).
\label{50}
\end{equation}%
As is seen from this formula in the absence of synchronism, there is a weak
dependence of radiation field on harmonics' number.

In the case of synchronism ($n_{s}=$ $n_{0}$), Eq. (\ref{49}) becomes 
\begin{equation}
\frac{\partial E_{s}}{\partial x}+\frac{n_{0}}{c}\frac{\partial E_{s}}{%
\partial t}=i^{s-1}\mu \frac{2\pi e\rho _{0}}{n_{0}\left( n_{0}^{2}-1\right)
^{1/2}}J_{s}(\alpha ).  \label{51}
\end{equation}

For the first harmonic (fundamental coherent radiation) the results repeat
almost exactly the case of wave circular polarization (Eqs. (\ref{40})--(\ref%
{42})), the only difference being that the coherence length in this case is $%
\sqrt{2}l_{c}$.

To determine the radiation on the other harmonics in the case of synchronism
consider the problem in the given field. Then, for large $x$ when 
\begin{equation*}
\frac{e\left( n_{0}^{2}-1\right) E_{0}x}{mc^{2}}\gg 1
\end{equation*}%
for the harmonics' amplitudes we have 
\begin{equation}
E_{s}=i^{s-1}\mu \frac{2\pi mc^{2}\rho _{0}}{n_{0}\left( n_{0}^{2}-1\right)
^{3/2}}\frac{1}{E_{0}}.  \label{52}
\end{equation}%
Hence, the radiation intensity on the harmonics 
\begin{equation}
I_{s}=\frac{c}{8\pi }|E_{s}|^{2}\simeq e^{2}c\frac{(\lambda _{0}^{3}\rho
_{0})^{2}}{\lambda _{0}^{4}}\left( \frac{\mathcal{E}_{0}}{mc^{2}}\right)
^{2}.  \label{53}
\end{equation}

As in the linear regime the coherence length increases as energy squared,
and the losses of the particles in the medium depend on energy
logarithmically, then the energy increase for amplification of weak signals
does not give an essential advantage. The optimal energy is $\mathcal{E}%
_{0}\sim mc^{2}$. Then $l_{c}\sim (r_{0}\lambda _{0}\rho _{0})^{-1}$, where $%
r_{0}=e^{2}/mc^{2}$ is the electron classical radius. The estimations show
that for the amplification of optical radiation in the capture regime with $%
n_{0}=\mathrm{const}$, electron beams of large densities are necessary. The
situation considerably will be improved if media with varying refraction
index $n_{0}(x)$ are used. Then along the direction of increase of $n_{0}(x)$
the particles will be continuously decelerated, and the wave continuously
amplified (a regime inverse to Cherenkov accelerator).

\section{Quantum Effects in Induced Cherenkov Process}

\subsection{Quantum Description of \textquotedblleft
Reflection\textquotedblright\ Phenomenon. Particle Beam Quantum Modulation
at X-Ray Frequencies}

Though the phenomenon of particle \textquotedblleft
reflection\textquotedblright\ from the front of a plane EM wave is of
classical nature, which means that quantum effects of above-barrier
reflection and tunnel passage should be small enough, nevertheless the
quantum consideration of this phenomenon is worthy of note in relation to
the appearance of an important coherent quantum effect as a result of
classical \textquotedblleft reflection\textquotedblright\ of particles. The
influence of spin interaction is not essential here; on the other hand, it
is quantitatively small enough in the induced Cherenkov process (for optical
frequencies) and may be neglected. The qualitative aspect of spin effects in
the induced Cherenkov process will be considered below. Neglecting the spin
interaction, the Dirac equation in quadratic form becomes the Klein--Gordon
equation, the exact solution of which can be obtained when the particle
initial velocity is parallel to the wave propagation direction ($\mathbf{p}%
_{\bot 0}=0$) and the latter is monochromatic of circular polarization ($%
\mathbf{A}^{2}\mathbf{(}\tau \mathbf{)=}\mathrm{const}$). To calculate the
probability of reflection from the wave barrier one needs to consider an EM
pulse with the envelope of intensity damped asymptotically at infinity. For
a laser pulse of form 
\begin{equation}
\xi ^{2}\left( \tau \right) =\frac{\xi _{0}^{2}}{\cosh ^{2}\frac{\tau }{\tau
_{0}}},  \label{84a}
\end{equation}%
($\xi _{0}^{2}$ is the maximal value of intensity and $\tau _{0}$ is the
half-width of the pulse) at $\xi _{0}>\xi _{cr}$ for the coefficient of
reflection we have 
\begin{equation}
R=\frac{\exp \left[ \pi \widetilde{\Omega }\tau _{0}\left( \frac{\xi _{0}}{%
\xi _{cr}}-1\right) \right] }{1+\exp \left[ \pi \widetilde{\Omega }\tau
_{0}\left( \frac{\xi _{0}}{\xi _{cr}}-1\right) \right] }.  \label{85}
\end{equation}%
where%
\begin{equation}
\widetilde{\Omega }=2\frac{\mathcal{E}_{0}}{\hbar \left( n^{2}-1\right) }%
\left\vert 1-n\frac{\mathrm{v}_{0}}{c}\right\vert  \label{85a}
\end{equation}%
is the quantum frequency corresponding to particle classical energy change
due to \textquotedblleft reflection\textquotedblright\ (see Eq. (\ref{7}))
and $\xi _{cr}$ is the classical value of critical intensity (\ref{6}). The
major quantity $\widetilde{\Omega }\tau _{0}$ in Eq. (\ref{85}) $\widetilde{%
\Omega }\tau _{0}\gg 1$ (for actual parameters of electron and laser beams
in a medium with refractive index $n-1\sim 10^{-4}$ the parameter $%
\widetilde{\Omega }\tau _{0}\sim 10^{15}\div 10^{11}$ for laser pulse
duration $\tau _{0}\sim 10^{-8}\div 10^{-12}$~$\mathrm{s}$), hence Eq. (\ref%
{85}) corresponds to above reflection regime $\xi _{0}>\xi _{cr}$.

This equation shows that $R=1$ with great accuracy (the coefficient of
tunnel passage in this case is of the order $\exp \left[ (-10^{15})\div
(-10^{11})\right] $). If $\xi _{0}<\xi _{cr}$ then the coefficient of
reflection $R=0$ with the same accuracy, i.e. the above barrier reflection
is negligibly small in this case. Thus, the quantum effects of tunnel
passage and above barrier reflection do not impact on the classical
phenomenon of particle \textquotedblleft reflection\textquotedblright\ from
the plane EM wave. This is physically clear since the Compton wavelength of
a particle (electron) is much smaller than the space size of actual EM
pulses. Nevertheless, due to the particle quantum feature as a result of
classical reflection the coherent effect of quantum modulation of the free
particle probability density and, consequently, electric current density
occurs because of superposition of an incident and reflected particle's
matter waves.

The density of electric current of the particle beam is modulated at
frequency $\widetilde{\Omega }$ 
\begin{equation}
\mathbf{J}(x,t)=\mathbf{J}_{0}\left\{ 1+\cos \left[ \widetilde{\Omega }%
\left( t-n\frac{x}{c}\right) -\varphi _{0}\right] \right\} ,  \label{88}
\end{equation}%
where $\mathbf{J}_{0}=\mathrm{const}$ is the electric current density of the
initially homogeneous and monochromatic particle beam. The modulation
frequency $\widetilde{\Omega }$ in actual cases lies in the X-ray domain as
follows from the estimation of particle classical energy change $\Delta 
\mathcal{E}$ due to \textquotedblleft reflection\textquotedblright\ ($%
\widetilde{\Omega }=$ $\Delta \mathcal{E}$ /$\hbar $). Note that quantum
modulation in contrast to classical modulation is exceptionally the feature
of a single particle and so is conserved after the interaction.

\subsection{Spin Effects in Induced Cherenkov Process}

Consider now the role of spin effects in the nonlinear quantum dynamics of a
spinor particle in the field of a plane monochromatic EM wave in a medium,
i.e., in the \textquotedblleft reflection\textquotedblright\ and capture
phenomenon of an electron. The exact solution of the Dirac equation for
external electron (with respect to wave) can be found for the
above-considered case when the particle initial velocity is parallel to the
wave propagation direction and the latter is of circular polarization ($%
\mathbf{A}(\tau )=\left\{ 0,A_{0}\sin \omega \tau ,A_{0}\cos \omega \tau
\right\} $). In the result of spin interaction two critical values of the
wave intensity appear corresponding to different initial spin projections $%
\sigma _{x}=\mp 1$ along the direction of particle motion: 
\begin{equation}
\xi _{cr1,2}^{2}=\left( \frac{\mathcal{E}_{0}}{mc^{2}}\right) ^{2}\frac{%
\left[ 1-n\frac{\mathrm{v}_{0}}{c}\pm \frac{\hbar \omega }{2\mathcal{E}_{0}}%
\left( n^{2}-1\right) \right] ^{2}}{\left( n^{2}-1\right) }.  \label{92}
\end{equation}%
and for particles energies we have respectively: 
\begin{equation}
\mathcal{E}_{1,2}=\mathcal{E}_{0}+\frac{\mathcal{E}_{0}}{n^{2}-1}\left( 1-n%
\frac{\mathrm{v}_{0}}{c}\pm \frac{\hbar \omega }{2\mathcal{E}_{0}}\left(
n^{2}-1\right) \right) \left( 1-\sqrt{1-\frac{\xi _{0}^{2}}{\xi _{cr1}^{2}}}%
\right) ,  \label{94}
\end{equation}%
For the reflected particles energies with the spin projections $\sigma
_{x}=+1$\ and $\sigma _{x}=-1$, respectively, we have: 
\begin{equation*}
\mathcal{E}_{3,4}=\mp \hbar \omega +\mathcal{E}_{0}+\frac{\mathcal{E}_{0}}{%
n^{2}-1}\left( 1-n\frac{\mathrm{v}_{0}}{c}\pm \frac{\hbar \omega }{2\mathcal{%
E}_{0}}\left( n^{2}-1\right) \right)
\end{equation*}%
\begin{equation}
\times \left( 1+\sqrt{1-\frac{\xi _{0}^{2}}{\xi _{cr1,2}^{2}}}\right) .
\label{97}
\end{equation}%
In particular, from this equation it follows that in Eq. (\ref{92}) $\xi
_{cr2}$ corresponds to a particle with the spin directed along the axis $OX$%
, while $\xi _{cr1}$ corresponds to the opposite one.

The expressions of particle wave functions show that the degeneration of
particle states over the spin projection that takes place in vacuum (Volkov
states) vanishes in a dielectriclike medium. In that case the wave function $%
\Psi _{1}$ corresponds to superposition state with energies $\mathcal{E}_{1}$
and $\mathcal{E}_{1}-\hbar \omega $, while $\Psi _{2}$ corresponds to
energies $\mathcal{E}_{2}$ and $\mathcal{E}_{2}+\hbar \omega $. The removal
of degeneration of Volkov states is related to the fact that in a medium
with refractive index $n>1$ in the intrinsic frame of reference of the wave
there is only a static magnetic field and the spin interaction with such a
field results in the splitting of the particle states as by the Zeeman
effect. The splitting value is: $\Delta \mathcal{E}=\left\vert \mathcal{E}%
_{1}-\mathcal{E}_{2}\right\vert =\left\vert \mathcal{E}_{4}-\mathcal{E}%
_{3}\right\vert $. In vacuum this splitting vanishes as it follows from Eqs.
(\ref{92}),(\ref{94}) and Dirac wave function in a medium passes to the
Volkov wave function (\cite{2006}).

The spin interaction in a medium within the nonlinear threshold phenomenon
of particle \textquotedblleft reflection\textquotedblright\ may lead to
particle beam polarization since the critical intensity (\ref{92}) depends
on spin projection along the direction of particle motion. Thus, if the
condition $\xi _{cr2}^{2}<$ $\xi ^{2}<\xi _{cr1}^{2}$ holds, then only the
particles with certain direction of the spin (along the axis $OX$) will be
reflected. Since the velocities of reflected particles are different from
the nonreflected ones, then by separating the particles after the
interaction a polarized beam may be obtained.

\subsection{Reflection of Electron From the Phase Lattice of Slowed EM Wave}

As was shown above, the exact solution of the Dirac equation can be achieved
only for the particular case when the particle initial velocity is parallel
to the wave propagation direction, which is monochromatic and is of circular
polarization. In other cases the quantum equations of motions (both
nonrelativistic and relativistic) are reduced to ordinary differential
equations of the second order of Hill or Mathieu type, the exact solution of
which are unknown. In these cases one needs to develop adequate
approximations for the quantum description of particle--wave nonlinear
interaction at the intensities close to critical value (indeed below the
threshold of classical \textquotedblleft reflection\textquotedblright\
phenomenon) when the probabilities of multiphoton absorption/emission become
maximal \cite{Q1,Q2,Q3}. The expected quantum effect in this case is
electron reflection from the phase lattice of slowed EM wave in a dielectric
medium \cite{Q1}. Note that on the base of this effect a nonlinear scheme of
coherent x-ray source (induced FEL) has been proposed \cite{Q4}.

To solve Dirac equation it is more straightforward to pass to the frame of
reference of the rest of the wave ($R$ frame moving with velocity $\mathrm{V}%
=c/n$). In the $R$ frame there is only the static magnetic field that will
be described by the following vector potential $\mathbf{A}_{R}=\left\{
0,A_{0}(x^{\prime })\cos k^{\prime }x^{\prime },0\right\} $, where the
wavenumber in this frame $k^{\prime }=\left( \omega /c\right) \sqrt{n^{2}-1}$%
. The wave function of a particle in the $R$ frame is connected with the
wave function in the laboratory frame $L$ by the Lorentz transformation of
the bispinors 
\begin{equation}
\Psi =\widehat{S}(\vartheta )\Psi _{R},\text{ }\widehat{S}(\vartheta )=ch%
\frac{\vartheta }{2}+\alpha _{x}sh\frac{\vartheta }{2}  \label{71}
\end{equation}%
($th\vartheta =\mathrm{V/}c=1/n$). Since the interaction Hamiltonian does
not depend on the time and transverse coordinates the eigenvalues of the
Hamiltonian $\widehat{H^{\prime }}$ and momentum operators $\widehat{p}%
_{y}^{\prime }$, $\widehat{p}_{z}^{\prime }$ are conserved: $\mathcal{E}%
^{\prime }=\mathrm{const}$, $p_{y}^{\prime }=\mathrm{const}$, $p_{z}^{\prime
}=\mathrm{const,}$the solution of Dirac equation for wave function $\Psi
_{R} $ can be represented in the form of a linear combination of free
solutions of the Dirac equation $\Psi _{i}^{(0)}$ with amplitudes $%
a_{i}(x^{\prime })$ depending only on $x^{\prime }$: 
\begin{equation}
\Psi _{R}(\mathbf{r}^{\prime }\mathbf{,}t^{\prime
})=\sum\limits_{i=1}^{4}a_{i}(x^{\prime })\Psi _{i}^{(0)}.  \label{72}
\end{equation}%
The solution in the form (\ref{72}) corresponds to the expansion of the wave
function in a complete set of the wave functions of a particle with certain
energy and transverse momentum $p_{y}^{\prime }$ (with longitudinal momenta $%
\pm (\mathcal{E}^{^{\prime }2}/c^{2}-p_{y}^{^{\prime }2}-m^{2}c^{2})^{1/2}$
and spin projections $S_{z}=\pm 1/2$). The latter are normalized to one
particle per unit volume. Since there is symmetry with respect to the
direction $\mathbf{A}_{R}$ (the $OY$ axis), we have taken, without loss of
generality, the vector $\mathbf{p}^{\prime }$ in the $XY$ plane $%
(p_{z}^{\prime }=0)$.

According to expansion (\ref{72}) the induced Cherenkov effect in the $R$
frame corresponds to elastic scattering process by which the reflection of
the particle from the wave field occurs: $p_{x}^{\prime }\rightarrow
-p_{x}^{\prime }$. However, in contrast to classical reflection when the
periodic wave field becomes a potential barrier for the particle at the
intensity $\xi >\xi _{cr}$, this quantum above-barrier reflection takes
place regardless of how weak the wave field is. Hence, the probability of
multiphoton absorption/radiation of the incident wave photons by the
particle in the $L$ frame, that is, induced Cherenkov effect, will be
determined by the probability of particle elastic reflection in the $R$
frame.

Substituting Eq. (\ref{72}) into Dirac equation and then multiplying by the
Hermitian conjugate functions we obtain a set of differential equations for
the unknown functions $a_{i}(x^{\prime })$. For simplicity we shall assume
that before the interaction there are only particles with specified
longitudinal momentum and spin state, i.e., 
\begin{equation}
\left\vert a_{1}(-\infty )\right\vert ^{2}=1,\ \left\vert a_{3}(+\infty
)\right\vert ^{2}=0,\ \left\vert a_{2}(-\infty )\right\vert ^{2}=0,\
\left\vert a_{4}(+\infty )\right\vert ^{2}=0.  \label{73a}
\end{equation}%
From the condition of conservation of the norm we have 
\begin{equation}
\left\vert a_{1}(x^{\prime })\right\vert ^{2}-\left\vert a_{3}(x^{\prime
})\right\vert ^{2}=\mathrm{const}  \label{73b}
\end{equation}%
and the probability of reflection is $\left\vert a_{3,4}(-\infty
)\right\vert ^{2}$.

The equations for amplitudes $a_{1}$, $a_{3}$ and $a_{2}$, $a_{4}$ are
separated and after unitarian transformation 
\begin{equation}
a_{1,3}(x^{\prime })=b_{1,3}(x^{\prime })\exp \left[ \pm \left( \frac{%
iep_{y}^{\prime }}{\hbar cp_{x}^{\prime }}\right) \int_{-\infty }^{x^{\prime
}}A_{y}(\eta )d\eta \mp i\frac{\vartheta ^{\prime }}{2}\right]  \label{74}
\end{equation}%
the problem simplifies and we obtain set of equations for the amplitudes $%
b_{1}(x^{\prime })$ and $b_{3}(x^{\prime })$: 
\begin{equation}
\frac{db_{1,3}(x^{\prime })}{dx^{\prime }}=-\sum\limits_{N=-\infty }^{\infty
}f_{N}\exp \left[ \mp \frac{i}{\hbar }(2p_{x}^{\prime }-N\hbar k^{\prime
})x^{\prime }\right] b_{3,1}(x^{\prime }),  \label{74a}
\end{equation}%
In unitarian transformation $\vartheta ^{\prime }$ is the angle between the
particle momentum and the direction of the wave propagation in the $R$
frame. The new amplitudes $b_{1}(x^{\prime })$ and $b_{3}(x^{\prime })$
satisfy the same initial conditions: $\left\vert b_{1}(-\infty )\right\vert
^{2}=1,$ $\left\vert b_{3}(+\infty )\right\vert ^{2}=0,$ according to Eq. (%
\ref{73a}). A similar set of equations is also obtained for the amplitudes $%
a_{2}(x^{\prime })$ and $a_{4}(x^{\prime })$, then after unitarian
transformation -for the amplitudes $b_{2}(x^{\prime })$ and $b_{4}(x^{\prime
})$.

The new amplitudes $b_{1}(x^{\prime })$ and $b_{3}(x^{\prime })$ satisfy the
same initial conditions: $\left\vert b_{1}(-\infty )\right\vert ^{2}=1,$ $%
\left\vert b_{3}(+\infty )\right\vert ^{2}=0,$ according to Eq. (\ref{73a})
and 
\begin{equation}
f_{N}=\frac{p^{\prime }}{2p_{y}^{\prime }}Nk^{\prime }J_{N}\left( 2\xi \frac{%
mc}{p_{x}^{\prime }}\frac{p_{y}^{\prime }}{\hbar k^{\prime }}\right) .\quad
\quad  \label{74d}
\end{equation}

Because of conservation of particle energy and transverse momentum (in $R$
frame) the real transitions in the field will occur from a $p_{x}^{\prime }$
state to the $-p_{x}^{\prime }$ one and, consequently, the probabilities of
multiphoton scattering will have maximal values for the resonant transitions 
\begin{equation}
2p_{x}^{\prime }=s\hbar k^{\prime }\qquad (s=\pm 1,\pm 2...).  \label{75}
\end{equation}%
The latter expresses the condition of exact resonance between the particle
de Broglie wave and the incident \textquotedblleft wave
lattice\textquotedblright . In the $L$ frame the inelastic scattering of the
particle on the moving phase lattice takes place and Eq. (\ref{75})
corresponds to the known Cherenkov conservation law 
\begin{equation}
\frac{2\mathcal{E}_{0}(1-n\frac{\mathrm{v}_{0}}{c}\cos \vartheta )}{(n^{2}-1)%
}=s\hbar \omega ,  \label{75a}
\end{equation}%
where $\vartheta $ is the angle between the particle momentum and the wave
propagation direction (the Cherenkov angle), and $\mathrm{v}_{0}$ and $%
\mathcal{E}_{0}$ are the particle initial velocity and energy in the $L$
frame.

At the exact resonance ($\delta _{s}=0$), according to the boundary
conditions (\ref{73b}) for the reflection coefficient we have 
\begin{equation}
R^{(s)}=\left\vert b_{3}^{(s)}(-\infty )\right\vert ^{2}=\tanh ^{2}\left[
f_{s}\triangle x^{\prime }\right] ,  \label{77a}
\end{equation}%
where $\triangle x^{\prime }$ is the coherent interaction length. The
reflection coefficient in the laboratory frame of reference is the
probability of absorption at $\mathrm{v}_{0}<c/n$ or emission at $\mathrm{v}%
_{0}>c/n$. The latter can be obtained expressing the quantities $f_{s}$ and $%
\triangle x^{\prime }$ by the quantities in this frame since the reflection
coefficient is Lorentz invariant. So 
\begin{equation}
R^{(s)}=\tanh ^{2}\left[ F_{s}\triangle \tau \right] ,  \label{77b}
\end{equation}%
where 
\begin{equation*}
F_{s}=\left[ \frac{(1-n\frac{\mathrm{v}_{0}}{c}\cos \vartheta )^{2}}{n^{2}-1}%
+\frac{\mathrm{v}_{0}^{2}}{c^{2}}\sin ^{2}\vartheta \right] ^{1/2}
\end{equation*}%
\begin{equation}
\times \frac{s\omega c}{2\mathrm{v}_{0}\sin \vartheta }J_{s}\left( \xi \frac{%
2m\mathrm{v}_{0}c\sin \vartheta }{\hbar \omega (1-n\frac{\mathrm{v}_{0}}{c}%
\cos \vartheta )}\right) \quad \quad   \label{77c}
\end{equation}%
and $\triangle \tau $ for actual cases is the laser pulse duration in the $L$
frame. The condition of applicability of resonant approximation is
equivalent to the condition $\left\vert F_{s}\right\vert \ll \omega $ which
restricts the intensity of the wave as well as the Cherenkov angle. Besides,
we must take into account the very sensitivity of the parameter $F_{s}$
toward the argument of Bessel function, according to Eq. (\ref{77c}). For
the wave intensities when $F_{s}\triangle \tau \gtrsim 1$ the reflection
coefficient is of the order of one that can occur for a large number of
photons $s\gg 1$ for the argument of the Bessel function $\alpha \sim s\gg 1$
in Eq. (\ref{77c}) (according to the asymptotic behavior of Bessel function $%
J_{s}(\alpha )$ at $\alpha \simeq s\gg 1$).

For the off resonant solution, when $\delta _{s}\neq 0$, but $%
f_{s}^{2}>\delta _{s}^{2}/4$ we have the following expression for the
reflection coefficient: 
\begin{equation}
R^{(s)}=\frac{f_{s}^{2}}{\Omega _{s}^{2}}\frac{\sinh ^{2}[\Omega
_{s}\triangle x^{\prime }]}{1+\frac{f_{s}^{2}}{\Omega _{s}^{2}}\sinh
^{2}[\Omega _{s}\triangle x^{\prime }]};\text{ \quad }\Omega _{s}=\sqrt{%
f_{s}^{2}-\delta _{s}^{2}/4},  \label{78}
\end{equation}%
which has the same behavior as in the case of exact resonance. In the
opposite case when $f_{s}^{2}\leq \delta _{s}^{2}/4$ the reflection
coefficient is an oscillating function of interaction length.

\subsection{Quantum Description of Capture Phenomenon in Induced Cherenkov
Process}

The multiphoton induced Cherenkov interaction in the capture regime
corresponding to transitions between the particle bound states occurs at the
nonzero initial angles of particle motion with respect to the wave
propagation direction, at which, as mentioned above, the Dirac or
Klein--Gordon equations are of Hill or Mathieu type and unable to solve it
exactly. However, as was shown in the quantum description of
\textquotedblleft reflection\textquotedblright\ phenomenon (free--free
transitions), the interaction at the arbitrary initial angle resonantly
connects two states of the particle (in the intrinsic frame of reference of
the wave the states with longitudinal momenta $p_{x}$ of the incident
particle and $p_{x}+s\hbar k$ of the scattered particle; $s$ is the number
of absorbed or radiated photons with a wave vector $\mathbf{k}$), which
makes available the application of resonant approximation to determine the
multiphoton probabilities of free--free transitions in induced nonlinear
Cherenkov process. Concerning the quantum description of the particle's
bound states in the capture regime one must take into account the
degeneration of initial states of free particles in the \textquotedblleft
longitudinal momentum\textquotedblright . Therefore, regardless of how weak
the field of the wave is, the usual perturbation theory in stimulated
Cherenkov process is not applicable because of such degeneration of the
states, and the interaction near the resonance is needed for description by
the secular equation. The latter, in particular, reveals the zone structure
of the particle states in the field of a transverse EM wave in a
dielectriclike medium. Note that in contrast to the zone structure for the
energy of electron states in a crystal lattice, the zone structure in this
process holds for the conserved quantity%
\begin{equation}
p_{\eta }=\frac{1}{2}\left( \frac{c}{n}p_{x}-\mathcal{E}\right) =\mathrm{%
const},  \label{80}
\end{equation}
as the energy could not be quantum characteristic of the state in the
nonstationary field of the wave \cite{Q1,Q2,Q3}.

At first we will consider the case of scalar particles. According to
Floquet's theorem the solution of Klein--Gordon equation in the wave
coordinate $\tau $ may be represented in the form 
\begin{equation}
U(\tau )=e^{i\frac{p_{\tau }}{\hbar }\tau }\sum\limits_{s=-\infty }^{\infty
}\Phi _{s}e^{-is\omega \tau },  \label{100}
\end{equation}%
where 
\begin{equation}
p_{\tau }^{2}\equiv \frac{\mathcal{E}_{0}^{2}}{(n^{2}-1)^{2}}\left[ \left(
1-n\frac{\mathrm{v}_{0}}{c}\cos \vartheta \right) ^{2}-(n^{2}-1)\left( \frac{%
mc^{2}}{\mathcal{E}_{0}}\right) ^{2}\xi _{0}^{2}\right]  \label{100a}
\end{equation}%
is the major quantity in the induced nonlinear Cherenkov process, which is
the renormalized (because of intensity effect) generalized momentum of the
particle in the laboratory frame conjugate to wave coordinate $\tau $. It
connects the \textquotedblleft width of initial Cherenkov
resonance\textquotedblright\ $1-$ $n\mathrm{v}_{0}/c$ and wave intensity ($%
\xi _{0}^{2}$) as the main relation between the physical quantities of this
process determining also the condition of nonlinear resonance ($\mathrm{v}%
_{x}(\xi )\mid _{\xi =\xi _{cr}}=c/n$). In the intrinsic frame of reference
of the wave $p_{\tau }$ corresponds to longitudinal momentum $p_{x}$ of the
particle on which the degeneration exists. The recurrent equation for the
coefficients $\Phi _{s}$ can be solved in approximation of the perturbation
theory by the wave function which is valid at the satisfaction of condition%
\begin{equation}
\left\vert s^{2}\hbar ^{2}\omega ^{2}-2s\hbar \omega p_{\tau }\right\vert
\gg \left\vert \frac{mc^{3}}{(n^{2}-1)}p_{0}\xi _{0}\sin \vartheta
\right\vert .  \label{101a}
\end{equation}

Regarding those values $p_{\tau }$ for which condition (\ref{101a}) does not
hold, the usual perturbation theory is already not applicable. In
particular, if the expression on the left-hand side of this condition is
zero, i.e., at $s=0$ and $s=\ell $ ($\ell =1,2,3,...$ ), when 
\begin{equation}
2\frac{p_{\tau }}{\hbar }=\ell \omega \text{,}  \label{102}
\end{equation}
it is evident that we already have two states $\Phi _{0}$ and $\Phi _{\ell }$%
, which are degenerated in the \textquotedblleft longitudinal
momentum\textquotedblright\ $p_{\tau }$, since $p_{\tau }^{2}=(p_{\tau
}-\ell \hbar \omega )^{2}$. Because of this double degeneration in the state
parameter $p_{\tau }$ for the definite $p_{\eta }$ of the initial
unperturbed system it is necessary to use perturbation theory for the
degenerated states on the basis of the secular equation. Thus, within
secular perturbation theory from the compatibility of equations for $\Phi
_{0}$ and $\Phi _{\ell }$ we have $\Delta _{\tau }=\pm \alpha _{1}$, where $%
\Delta _{\tau }$ is the correction to the value $p_{\tau }^{2}$ at the
fulfillment of condition (\ref{102}) for $\ell =1$: 
\begin{equation}
\Delta _{\tau }\equiv \frac{8n^{2}p_{\eta }^{(0)}}{(n^{2}-1)^{2}}p_{\eta
}^{(1)};\qquad \alpha _{1}\equiv \frac{mc^{3}p_{0}\xi _{0}\sin \vartheta }{%
(n^{2}-1)}.  \label{104a}
\end{equation}%
By the standard method from Eq. (\ref{104a}) one can obtain the following
set of equations for the amplitudes $\Phi _{0}$ and $\Phi _{1}$:. The signs
\textquotedblleft $+$\textquotedblright\ and \textquotedblleft $-$%
\textquotedblright\ relate to $p_{\tau }^{2}>\hbar ^{2}\omega ^{2}/4$ and $%
0< $ $p_{\tau }^{2}<\hbar ^{2}\omega ^{2}/4$, respectively. Thus, at the
fulfillment of condition (\ref{102}) we have a jump in the value of $p_{\tau
}^{2}$, which is equal to $2\alpha _{1}$, i.e., 

\begin{equation*}
\frac{\mathcal{E}_{0}^{2}}{(n^{2}-1)^{2}}\left\{ \left( 1-n\frac{\mathrm{v}%
_{0}}{c}\cos \vartheta \right) ^{2}-(n^{2}-1)\left( \frac{mc^{2}}{\mathcal{E}%
_{0}}\right) ^{2}\xi _{0}^{2}\right\} 
\end{equation*}%
\begin{equation*}
\geq \frac{\hbar ^{2}\omega ^{2}}{4}+\alpha _{1},
\end{equation*}%
\begin{equation*}
0\leq \frac{\mathcal{E}_{0}^{2}}{(n^{2}-1)^{2}}\left\{ \left( 1-n\frac{%
\mathrm{v}_{0}}{c}\cos \vartheta \right) ^{2}-(n^{2}-1)\left( \frac{mc^{2}}{%
\mathcal{E}_{0}}\right) ^{2}\xi _{0}^{2}\right\} 
\end{equation*}%
\begin{equation}
\leq \frac{\hbar ^{2}\omega ^{2}}{4}-\alpha _{1}.  \label{106}
\end{equation}

For $\ell =1$ the matrix element of transition from state $\Phi _{0}$ to
state $\Phi _{1}$ (here we note the state without a phase) is equal to $%
\alpha _{1}$. For large $\ell $ ($\ell \geq 2$) the matrix element of
transition $\Phi _{0}\longleftrightarrow \Phi _{\ell }$ is equal to zero in
the first order of perturbation theory. In this case it makes sense to take
into account the transitions to the states with other energies in higher
order. For example, for $\ell =2$ it is necessary to consider the
transitions $\Phi _{0}\rightarrow \Phi _{1}$ and $\Phi _{0}\rightarrow \Phi
_{2}$. For arbitrary $\ell $ the matrix element of transition is defined by 
\begin{equation}
\alpha _{\ell }=\frac{\alpha _{1}^{\ell }}{\left( (\ell -1)!\right)
^{2}\left( \hbar \omega \right) ^{2(\ell -1)}}.  \label{107}
\end{equation}%
It should be noted that here it is also necessary to take into account the
corrections to the energy eigenvalue of state $\Phi _{0}$ in the appropriate
order, however, the latter are only of quantitative character, unlike the
qualitative corrections (\ref{107}), and will be omitted.

As is seen from Eq. (\ref{106}), the permitted and forbidden zones arise for
the particle states in the wave. The widths of permitted zones in the
general case of $\ell $-photon resonance are defined from the condition 

\begin{equation*}
\frac{\ell ^{2}\hbar ^{2}\omega ^{2}}{4}+\alpha _{\ell }
\end{equation*}%
\begin{equation*}
\leq \frac{\mathcal{E}_{0}^{2}}{(n^{2}-1)^{2}}\left\{ \left( 1-n\frac{%
\mathrm{v}_{0}}{c}\cos \vartheta \right) ^{2}-(n^{2}-1)\left( \frac{mc^{2}}{%
\mathcal{E}_{0}}\right) ^{2}\xi _{0}^{2}\right\} 
\end{equation*}%
\begin{equation}
\leq \frac{(\ell +1)^{2}\hbar ^{2}\omega ^{2}}{4}-\alpha _{\ell +1}.
\label{107a}
\end{equation}

Such zone structure for the particle states in the wave arises in
dielectriclike media because of particle capture by the wave and periodic
character of the field -- quantum influence of infinite \textquotedblleft
potential\textquotedblright\ wells on the particle states similar to zone
structure of electron states in a crystal lattice \cite{1913}.

Consider now the case of spinor particles. Proceeding from the Dirac
equation for the components of the spinor $V$: 
\begin{equation}
V_{1}(\tau )=e^{i\frac{p_{\tau }}{\hbar }\tau }\sum\limits_{s=-\infty
}^{\infty }K_{s}e^{-is\omega \tau }.  \label{116}
\end{equation}

Repeating the procedure as in the case of scalar particles, we obtain the
Bragg condition (\ref{102}), at which it is necessary to use the secular
perturbation theory for degenerated states. At $\ell =1$ we obtain the
following system of equations for coefficients $K_{0}$ and $K_{1}$: 
\begin{equation}
\left\{ 
\begin{array}{c}
\Delta _{\tau }K_{0}+\left( -i\alpha _{1}+\frac{1}{2}\mu \right) K_{1}=0, \\ 
\left( i\alpha _{1}+\frac{1}{2}\mu \right) K_{0}+\Delta _{\tau }K_{1}=0,%
\end{array}%
\right.  \label{117}
\end{equation}%
where 
\begin{equation}
\mu =\frac{\hbar ecH}{n\sqrt{n^{2}-1}}.  \label{117a}
\end{equation}%
From Eq. (\ref{117}) for the correction to $p_{\tau }^{2}$ we obtain 
\begin{equation}
\Delta _{\tau }=\pm \left( \frac{1}{4}\mu ^{2}+\alpha _{1}^{2}\right) ^{%
\frac{1}{2}}.  \label{117b}
\end{equation}%
It is easy to see that $K_{1}=\mp K_{0}e^{i\varphi }$, where $tg\varphi
=2\alpha _{1}/\mu $. Hence, each spinor component of particle wave function
has two values corresponding to the top and bottom borders of the first
forbidden zone: 
\begin{equation}
V_{1}^{\pm }\left( \tau \right) =K_{0}\left( e^{i\frac{\omega }{2}\tau }\mp
e^{-i\frac{\omega }{2}\tau +i\varphi }\right) ,  \label{118}
\end{equation}%
For $V_{2}\left( \tau \right) $ we have the same expressions as (\ref{118}),
where it is only necessary to replace $\varphi $ by $-\varphi $.

At $\ell =2$ we have already two channels for the transition from state $%
K_{0}$ to state $K_{2}$. The first is the result of the interaction
described by a term quadratic in the field ($\sim A^{2}$), the matrix
element of which at $\ell =2$ is equal to $\left( mc^{2}\right) ^{2}\xi
_{0}^{2}/4\hbar ^{2}(n^{2}-1)$, and the second channel proceeds both in the
case of scalar particles via transitions $K_{0}\rightarrow K_{1}$ and $%
K_{0}\rightarrow K_{2}$ , stipulated by the charge interaction $\sim \mathbf{%
pA}$, as well as for the spin interaction, the matrix elements of which at
each transition are equal to $-i\alpha _{1}$ and $\mu /2$, respectively.
Therefore, for two-photon transition for the main parameter -- correction to 
$p_{\tau }^{2}$ we obtain 
\begin{equation}
\Delta _{\tau }=\pm \frac{1}{\hbar ^{2}\omega ^{2}}\left[ \left( \frac{1}{4}%
\mu ^{2}-\alpha _{1}^{2}+\frac{\hbar ^{2}\omega ^{2}(mc^{2})^{2}\xi _{0}^{2}%
}{4(n^{2}-1)}\right) ^{2}+\alpha _{1}^{2}\mu ^{2}\right] ^{\frac{1}{2}}
\label{119}
\end{equation}%
The obtained results for spinor particles are valid at the fulfillment of
the condition $\left\vert \Delta _{\tau }\right\vert \ll \hbar ^{2}\omega
^{2}/4$.

Thus, the quantum picture of induced Cherenkov interaction for charged
spinor particles does not differ qualitatively from the case of scalar
particles, i.e., the spin interaction results only in quantitative
corrections to the quantities describing the process. However, in the
absence of charge interaction ($\mathbf{pA}=0$) in the first order in the
field, i.e., for one-photon interaction, the first forbidden zone ($\ell =1$%
) does not exist for scalar particles, but exists for spinor particles due
to the spin interaction.

\subsection{Quantum Modulation of Charged Particles at Optical Harmonics}

Coherent interaction of charged particles with a plane EM wave of intensity
smaller than the critical one in the induced Cherenkov process leads to
quantum modulation of the particle probability density and, consequently,
current density after the interaction at the wave fundamental frequency and
its harmonics. In contrast to classical modulation of particles' current
density proceeding in the free drift region after the interaction (see, e.g. 
\cite{P1,P2}) and conserving for short distances, the quantum modulation,
being quantum feature of a single particle, is conserved after the
interaction unlimitedly long \cite{mod}. To reveal this quantum coherent
effect it is necessary to take into account the quantum character of
particle--wave interaction entirely, i.e. the quantum recoil as well, in
contrast to the eikonal approximation for particle wave function at the
description of multiphoton interaction with strong EM wave. The mathematical
point of view requires taking into account in quantum equations of motion
the second-order derivatives of the wave function as well, which are
neglected in the eikonal approximation (see, below the description of the
diffraction effect).

To describe the effect of particle quantum modulation with regard to the
wave harmonics we will solve the Klein--Gordon equation by perturbation
theory in the field of monochromatic wave ($\mathbf{A}(\tau )=\left\{
0,A_{0}\cos \omega \tau ,A_{0}\sin \omega \tau \right\} $) of intensity $\xi
_{0}<\xi _{cr}\ll 1$ at which one can neglect again the constant term $\sim
A_{0}^{2}$. Then looking for the solution of Klein--Gordon equation in the
form which will be solved in the approximation of perturbation theory by
wave function. Thus, the amplitude of the particle wave function
corresponding to $s$-photon induced radiation ($s>0$) and absorption,
respectively, reads 
\begin{equation}
\Psi _{\pm s}=\frac{\left( \pm 1\right) ^{s}}{s!}\frac{b^{s}}{\left( \mu \pm
\bigtriangleup _{\hbar }\right) \left( \mu \pm 2\bigtriangleup _{\hbar
}\right) \cdots \left( \mu \pm s\bigtriangleup _{\hbar }\right) },
\label{3.105}
\end{equation}%
Here the dimensionless parameter of one-photon interaction 
\begin{equation}
b=\frac{1}{2}\frac{eA_{0}}{\hbar \omega }\frac{\mathrm{v}_{0}}{c}\sin
\vartheta _{0}  \label{3.107}
\end{equation}%
is the small parameter of perturbation theory: $\left\vert b\right\vert $ $%
\ll 1$ and 
\begin{equation}
\mu =1-n\frac{\mathrm{v}_{0}}{c}\cos \vartheta _{0};\qquad \bigtriangleup
_{\hbar }=\left( n^{2}-1\right) \frac{\hbar \omega }{2\mathcal{E}_{0}}
\label{3.107'}
\end{equation}%
are the dimensionless Cherenkov resonance width and quantum recoil
parameter, respectively.

The current density of the particles after the interaction is given by the
formula:%
\begin{equation*}
\mathbf{j}(t,x)=\mathbf{j}_{0}\left\{ 1+2\sum_{s=1}^{\infty }\frac{b^{s}}{s!}%
\left[ \frac{1}{\left( \mu +\Delta _{\hbar }\right) \cdots \left( \mu
+s\Delta _{\hbar }\right) }\right. \right.
\end{equation*}%
\begin{equation*}
\left. +\frac{(-1)^{s}}{\left( \mu -\Delta _{\hbar }\right) \cdots \left(
\mu -s\Delta _{\hbar }\right) }\right] \cos s\omega \left( t-nx/c\right)
\end{equation*}%
\begin{equation*}
+2\sum_{s=1}^{\infty }\sum_{s^{\prime }=1}^{\infty }(-1)^{s^{\prime }}\frac{%
b^{s+s^{\prime }}}{s!s^{\prime }!}\cos \left[ \left( s+s^{\prime }\right)
\omega \left( t-nx/c\right) \right]
\end{equation*}%
\begin{equation}
\left. \times \frac{1}{\left( \mu +\Delta _{\hbar }\right) \cdots \left( \mu
+s\Delta _{\hbar }\right) \cdot \left( \mu -\Delta _{\hbar }\right) \cdots
\left( \mu -s^{\prime }\Delta _{\hbar }\right) }\right\} ,  \label{3.109}
\end{equation}%
where $\mathbf{j}_{0}=\mathrm{const}$ is the current density of initially
uniform particle beam. As is seen from Eq. (\ref{3.109}) as a result of
direct and inverse induced Cherenkov effect the current density of initially
uniform particle beam is modulated at the wave fundamental frequency and its
harmonics. This is a result of coherent superposition of particle states
with various energy and momentum due to absorbed and emitted photons in the
radiation field that remains after the interaction unlimitedly long (for a
monochromatic beam).\ 

The denominators in Eq. (\ref{3.109}) becomes zero at the fulfillment of
exact quantum conservation law for multiphoton Cherenkov process (\ref{75a}%
). In this case perturbation theory is not applicable and the consideration
in the scope of above-developed secular perturbation is required. However,
in actual cases because of nonmonochromaticity of particle beams the width
of Cherenkov resonance is rather larger than quantum recoil ($\Delta _{\hbar
}\ll \mu $) and one can neglect the latter in Eq. (\ref{3.109}). Then the
modulation depth ($B$) at the wave fundamental frequency is expressed via
critical intensity (\ref{9}): $B=\xi /2\xi _{cr}(\vartheta )$. The latter
shows that the effect of quantum modulation at the stimulating wave
harmonics proceeds at intensities smaller than the critical one when the
induced Cherenkov interaction of the particles with the periodic wave field
(photons) occurs. In the opposite case the interaction proceeds with the
potential barrier, and as was shown above the quantum modulation of the
particles due to \textquotedblleft reflection\textquotedblright\ phenomenon
occurs because of interference of the incident and reflected electron's
matter waves. For this reason the modulation frequency (actually x-ray)
corresponds to particle's energy exchange in the result of interaction with
the moving barrier. On the other hand, it is clear that the modulated by
anyway particle beam is a coherent source of EM radiation.

\section{Vacuum Versions of \textquotedblleft Reflection\textquotedblright\
and Capture Phenomenon}

\subsection{Induced Nonlinear Compton Process}

Consider now the induced Compton and undulator processes in vacuum, at which
the restriction on the wave intensity taking place in dielectric media for
induced Cherenkov process vanishes and one can use laser pulses of extremely
large intensities to achieve laser acceleration of particles of superhigh
energies by considering phenomena, as well as realization of many nonlinear
QED phenomena from vacuum. At first let investigate the classical dynamics
of a charged particle at the interaction with two bichromatic
counterpropagating (along the axis $OX$) plane EM waves (with frequencies $%
\omega _{1}$ and $\omega _{2}$) in vacuum. Relativistic equations of motion
allow exact solution in case of monochromatic waves of circular polarization
and if the particle initial momentum is directed along the axis of waves'
propagation (initial transverse momentum $\mathbf{P}_{0\perp }=0$). Then for
the particle energy in the field we have \cite{Comp._Reflec}:

\begin{equation*}
\mathcal{E}=\frac{\mathcal{E}_{0}}{n_{1}^{2}-1}\left\{ n_{1}^{2}\left( 1-%
\frac{\mathrm{v}_{0}}{cn_{1}}\right) \mp \left[ \left( 1-n_{1}\frac{\mathrm{v%
}_{0}}{c}\right) ^{2}-\left( n_{1}^{2}-1\right) \left( \frac{mc^{2}}{%
\mathcal{E}_{0}}\right) ^{2}\right. \right.
\end{equation*}%
\begin{equation}
\left. \times \left[ \xi _{1}^{2}+\xi _{2}^{2}+2\xi _{1}\xi _{2}\cos \left(
\omega _{1}-\omega _{2}\right) \left( t-n_{1}\frac{x}{c}\right) \right] %
\right] ^{1/2}\Biggr\}.  \label{5.7}
\end{equation}%
The parameter $n_{1}$ included in Eq. (\ref{5.7}) is 
\begin{equation}
n_{1}=\frac{\omega _{1}+\omega _{2}}{\left\vert \omega _{1}-\omega
_{2}\right\vert }  \label{5.8}
\end{equation}%
and the parameters $\xi _{1,2}\equiv eE_{1,2}/mc\omega _{1,2}$ (the waves
are turned on and turned off adiabatically at $t\rightarrow \mp \infty $).

As is seen from Eq. (\ref{5.7}) due to the effective interaction of the
particle with the counterpropagating waves a slowed traveling wave in vacuum
arises. The parameter $n_{1}$ denotes the refractive index of this
interference wave and since $n_{1}>1$ (see Eq. (\ref{5.8})) the phase
velocity of the effective traveling wave $\mathrm{v}_{ph}=c/n_{1}<c$. Then
the expression under the root in Eq. (\ref{5.7}) evidences the peculiarity
in the interaction dynamics like the induced Cherenkov one that causes the
analogous threshold phenomena of particle \textquotedblleft
reflection\textquotedblright\ and capture by the interference wave in the
induced Compton process. Hence, omitting the same procedure related to
bypass of the multivalence and complexity of Eq. (\ref{5.7}), which has been
made in detail for the analogous expression in the Cherenkov process, we
will present the final results for particle \textquotedblleft
reflection\textquotedblright\ and capture by the effective interference wave
in the induced Compton process. The threshold value of the \textquotedblleft
reflection\textquotedblright\ phenomenon or the critical field for nonlinear
Compton resonance is 
\begin{equation}
\xi _{cr}\left( \omega _{1,2}\right) \equiv \left( \xi _{1}+\xi _{2}\right)
_{cr}=\frac{\mathcal{E}_{0}}{mc^{2}}\frac{\left\vert \omega _{1}\left( 1-%
\frac{\mathrm{v}_{0}}{c}\right) -\omega _{2}\left( 1+\frac{\mathrm{v}_{0}}{c}%
\right) \right\vert }{2\sqrt{\omega _{1}\omega _{2}}}.  \label{5.9}
\end{equation}%
If one knows the longitudinal velocity $\mathrm{v}_{x}$ of the particle in
the field, then it is easy to see that $\xi _{cr}\left( \omega _{1,2}\right) 
$ is the value of the total intensity of counterpropagating waves at which $%
\mathrm{v}_{x}$ becomes equal to the phase velocity of the effective
interference wave: $\mathrm{v}_{x}=\mathrm{v}_{ph}=c/n_{1}$ irrespective of
the magnitude of particle initial velocity $\mathrm{v}_{0}$. The latter is
the condition of coherency of induced Compton process 
\begin{equation}
\omega _{1}\left( 1-\frac{\mathrm{v}_{x}}{c}\right) =\omega _{2}\left( 1+%
\frac{\mathrm{v}_{x}}{c}\right) .  \label{5.10}
\end{equation}%
Under condition (\ref{5.10}) the nonlinear resonance in the field of
counterpropagating waves of different frequencies occurs and because of
induced Compton radiation/absorption the particle velocity becomes smaller
or larger than the phase velocity of the interference wave and the particle
leaves the slowed effective wave. In the rest frame of the latter the
particle swoops on the motionless barrier (if $\xi _{1}+\xi _{2}>\xi
_{cr}\left( \omega _{1,2}\right) $) and the elastic reflection occurs. In
the laboratory frame it corresponds to inelastic \textquotedblleft
reflection\textquotedblright\ and from Eq. (\ref{5.7}) for particle energy
after the \textquotedblleft reflection\textquotedblright\ ($\xi
_{1,2}\rightarrow 0$ adiabatically at $t\rightarrow +\infty $) we have 
\begin{equation}
\mathcal{E=E}_{0}\frac{\omega _{1}^{2}\left( 1-\frac{\mathrm{v}_{0}}{c}%
\right) +\omega _{2}^{2}\left( 1+\frac{\mathrm{v}_{0}}{c}\right) }{2\omega
_{1}\omega _{2}}.  \label{5.11}
\end{equation}%
From this equation it follows that the energy of the particle with the
initial velocity $\mathrm{v}_{0}=c\left\vert \omega _{1}-\omega
_{2}\right\vert /\left( \omega _{1}+\omega _{2}\right) $ corresponding to
the resonance value of the induced Compton process does not change after the
interaction ($\mathcal{E}=\mathcal{E}_{0}$). For such particle $\xi
_{cr}\left( \omega _{1,2}\right) =0$, i.e., it cannot enter the field: $\xi
_{1}=\xi _{2}=0$. The particle with the initial velocity $\mathrm{v}%
_{0}>c\left\vert \omega _{1}-\omega _{2}\right\vert /\left( \omega
_{1}+\omega _{2}\right) $ after the \textquotedblleft
reflection\textquotedblright\ is decelerated, while at $\mathrm{v}%
_{0}<c\left\vert \omega _{1}-\omega _{2}\right\vert /\left( \omega
_{1}+\omega _{2}\right) $ it is accelerated because of direct and inverse
induced Compton processes. At the acceleration the particle absorbs photons
from the wave of frequency $\omega _{1}$ and coherently radiates into the
wave of frequency $\omega _{2}$ if $\omega _{1}>$ $\omega _{2}$ and at the
deceleration the inverse process takes place. Hence, at the particle
acceleration the amplification of the wave of a smaller frequency holds,
while at the deceleration the wave of a larger frequency is amplified.

In the case of $\omega _{1}=$ $\omega _{2}\equiv \omega $ the refractive
index of the interference wave $n_{1}=\infty $ and nonlinear interaction of
the particle with the strong standing wave occurs. It is evident that in
this case the process is elastic: $\mathcal{E}=\mathcal{E}_{0}=\mathrm{const}
$ (see Eq. (\ref{5.11})) and for the longitudinal momentum of the particle
in the field we have 
\begin{equation}
p_{x}=\pm \sqrt{p_{0}^{2}-m^{2}c^{2}\left( \xi _{1}^{2}+\xi _{2}^{2}+2\xi
_{1}\xi _{2}\cos \frac{2\omega }{c}x\right) }.  \label{5.12}
\end{equation}%
From this equation it is seen that at $\xi _{1}+\xi _{2}>$ $\xi _{cr}\left(
\omega \right) =\left\vert p_{0}\right\vert /mc$ the standing wave becomes a
potential barrier for the particle and elastic reflection occurs: the root
changes its sign and $p_{x}=-p_{0}$ (if $\xi _{1}+\xi _{2}<$ $\xi
_{cr}\left( \omega \right) $ we have $p_{x}=p_{0}$).

Because of limitation on the volume we will not consider here the capture
phenomenon in vacuum processes. The reader interested in those may find it
in Refs. \cite{2006,2016}.

\subsection{Induced Nonlinear Undulator/Wiggler Process}

Electrons "reflection" or capture phenomenon is also possible by a plane EM
wave propagating in electric or magnetic undulator/wiggler \cite{Und._Reflec}%
. As far as descriptions of electron dynamics in a plane monochromatic wave
in the electric and magnetic undulators are coincide in many features, here
we will consider only more important case of magnetic undulator/wiggler,
which is currently the most perspective coherent tool with extremely large
length of coherency, specifically due to which the x-ray free electron laser
has been realized in the wiggler \cite{Brau,Freund}.

Consider the nonlinear dynamics of a charged particle at the interaction
with a strong EM wave in a magnetic undulator/wiggler. Let a particle with
an initial velocity $\mathrm{v}_{0}=\mathrm{v}_{0x}$ enters into a magnetic
undulator with circularly polarized field 
\begin{equation}
\mathbf{H}(x)=\left\{ 0,-H\cos \frac{2\pi }{l}x,H\sin \frac{2\pi }{l}%
x\right\}  \label{5.18}
\end{equation}%
($l$ is the space period or step of an undulator) along the axis of which
propagates a plane monochromatic EM wave of circular polarization.
Relativistic equations of motion in this case allow exact solution and for
the particle energy we have \cite{Und._Reflec}: 
\begin{equation*}
\mathcal{E}=\frac{\mathcal{E}_{0}}{n_{2}^{2}-1}\Biggl\{n_{2}^{2}\left( 1-%
\frac{\mathrm{v}_{0}}{cn_{2}}\right) \mp \Biggl[\left( 1-n_{2}\frac{\mathrm{v%
}_{0}}{c}\right) ^{2}-\left( n_{2}^{2}-1\right) \left( \frac{mc^{2}}{%
\mathcal{E}_{0}}\right) ^{2}
\end{equation*}%
\begin{equation}
\times \left[ \xi _{0}^{2}+\xi _{H}^{2}-2\xi _{0}\xi _{H}\cos \omega (t-n_{2}%
\frac{x}{c})\right] \Biggr]^{1/2}\Biggr\}  \label{5.25}
\end{equation}%
where relativistic invariant dimensionless interaction parameter 
\begin{equation}
\xi _{H}=\frac{elH}{2\pi mc^{2}}  \label{5.26}
\end{equation}%
(for large magnitudes of undulator field strength $H$ and space period $l$
when $\xi _{H}>1$ such undulator is called a wiggler).

From Eq. (\ref{5.25}) it follows that at the particle--wave nonlinear
resonance interaction in the undulator an effective slowed traveling wave is
formed as in the induced Compton process. The parameter 
\begin{equation}
n_{2}=1+\frac{\lambda }{l}  \label{5.27}
\end{equation}%
is the refractive index of this slowed wave, which causes the analogous
threshold phenomenon of particle \textquotedblleft
reflection\textquotedblright\ in the induced undulator process. The
effective critical field at which the nonlinear resonance and then the
particle \textquotedblleft reflection\textquotedblright\ take place in the
undulator, is 
\begin{equation}
\xi _{cr}\left( \frac{\lambda }{l}\right) \equiv \left( \xi _{0}+\xi
_{H}\right) _{cr}=\frac{\left\vert 1-\left( 1+\frac{\lambda }{l}\right) 
\frac{\mathrm{v}_{0}}{c}\right\vert }{\sqrt{\frac{2\lambda }{l}\left( 1+%
\frac{\lambda }{2l}\right) }}\frac{\mathcal{E}_{0}}{mc^{2}}.  \label{5.28}
\end{equation}%
At this value of the resulting field the longitudinal velocity of the
particle $\mathrm{v}_{x}$ reaches the resonant value in the field at which
the condition of coherency in the undulator 
\begin{equation}
\frac{2\pi }{l}\mathrm{v}_{x}=\omega \left( 1-\frac{\mathrm{v}_{x}}{c}\right)
\label{5.29}
\end{equation}%
is satisfied. The latter has a simple physical explanation in the intrinsic
frame of the particle. In this frame of reference the static magnetic field (%
\ref{5.18}) becomes a traveling EM wave with the frequency 
\begin{equation*}
\omega ^{\prime }=2\pi \mathrm{v}_{x}/l\sqrt{1-\mathrm{v}_{x}^{2}/c^{2}}
\end{equation*}%
and phase velocity $\mathrm{v}_{ph}=\mathrm{v}_{x}$. For coherent
interaction process this frequency must coincide with the Doppler-shifted
frequency of stimulated wave.

The energy of the particle after the \textquotedblleft
reflection\textquotedblright\ (in Eq. (\ref{5.25}) $\xi _{0}=\xi _{H}=0$ at
the sign \textquotedblleft $+$\textquotedblright\ before the root) is 
\begin{equation}
\mathcal{E=E}_{0}\left[ 1+\frac{1-\left( \mathrm{v}_{0}/c\right) \left(
1+\lambda /l\right) }{\left( \lambda /l\right) \left( 1+\lambda /2l\right) }%
\right] .  \label{5.30}
\end{equation}%
From this equation it follows that the particle with the initial velocity $%
\mathrm{v}_{0}<c/(1+\lambda /l)$ after the \textquotedblleft
reflection\textquotedblright\ accelerates, while at $\mathrm{v}%
_{0}>c/(1+\lambda /l)$ it decelerates because of induced undulator radiation.

The \textquotedblleft reflection\textquotedblright\ phenomenon of charged
particles from a plane EM wave, as was shown in the induced Cherenkov
process, may be used for monochromatization of the particle beams. Note that
the considered vacuum versions of this phenomenon are more preferable for
this goal taking into account the influence of negative effects of the
multiple scattering and ionization losses in a medium. On the other hand,
the refractive index of the effective slowed waves in vacuum $n_{1}$ or $%
n_{2}$ in corresponding induced Compton and undulator processes may be
varied choosing the appropriate frequencies of counterpropagating waves or
wiggler step. In particular, for monochromatization of particle beams with
moderate or low energies via the induced Cherenkov process one needs a
refractive index of a medium $n-1\sim 1$ that corresponds to solid states.
Meanwhile, such values of effective refractive index may be reached in the
induced Compton process at the frequencies $\omega _{1}\sim \omega _{2}$ of
the counterpropagating waves. However, we will not consider here the
possibility of particle beam monochromatization on the basis of the vacuum
versions of \textquotedblleft reflection\textquotedblright\ phenomenon since
the principle of conversion of energetic or angular spreads is the same. To
study the subject in more detail we refer the reader to original papers \cite%
{Comp._mon.,und._mon.,conv.}.

Considered above quantum effects in the induced Cherenkov process as in the
above critical regime, as well as below the critical point also take place
in described vacuum versions of particle \textquotedblleft
reflection\textquotedblright\ and capture phenomena at the induced Compton
and undulator/wiggler nonlinear interaction. However, because of similar
physical picture and investigation methods we will not repeat here the
consideration of these effects.

\subsection{Inelastic Diffraction of Electron on a Traveling EM Wave}

As was mentioned in Introduction, at the wave intensities below the
threshold of particle \textquotedblleft reflection\textquotedblright\ and
capture phenomena, the inelastic diffraction scattering of electron matter
wave -- de Broglie wave on a traveling light phase-lattice in the induced
Cherenkov, Compton, and undulator processes is possible. For efficient
probabilities of multiphoton diffraction effect on such wave-gratings the
intensities of slowed traveling wave should be close to the critical value: $%
\xi \lesssim \xi _{cr}$ to be near the nonlinear resonance of particle-wave
interaction in each induced process.

We will proceed the solution of diffraction effect in general form for
Cherenkov and vacuum processes. As the coordinates $\mathbf{r}_{\bot
}=\left\{ y,z\right\} $ are cyclic, then the corresponding components of
generalized momentum $\mathbf{p}_{\bot }$ are conserved. Hence, according to
Floquet's theorem the solution of Klein--Gordon equation may be sought in
the form 
\begin{equation}
\Psi =e^{\frac{i}{\hbar }\mathbf{p}_{\bot }\mathbf{r}}\sum_{s}C_{s}\left(
t\right) e^{\frac{i}{\hbar }\left( p_{x}+s\hbar k\right) x}e^{-\frac{i}{%
\hbar }\left( \mathcal{E}+s\hbar k\mathrm{v}_{ph}\right) t}.  \label{Psi1}
\end{equation}%
where $\mathcal{E}=\sqrt{c^{2}\mathbf{p}^{2}+m_{\ast }^{2}c^{4}}$and we will
assume that $C_{s}\left( t\right) $ are slowly varying functions: $%
\left\vert \partial C_{s}/\partial t\right\vert \ll \mathcal{E}\left\vert
C_{s}\right\vert /\hbar ,$ and $\mathcal{E}\ll s\hbar k\mathrm{v}_{ph}$
(this condition is always satisfied for optical photons). From the
Klein--Gordon equation with Eq. (\ref{Psi1}) for the coefficients $%
C_{s}\left( t\right) $ we obtain the set of equations%
\begin{equation}
i\frac{\partial C_{s}\left( t\right) }{\partial t}+\Gamma _{s}C_{s}\left(
t\right) =\frac{\mathcal{W}_{eff}}{4\hbar \mathcal{E}}\left( C_{s-1}\left(
t\right) +C_{s+1}\left( t\right) \right) ,  \label{M2}
\end{equation}%
where%
\begin{equation*}
\Gamma _{s}=\frac{2\mathcal{E}s\hbar k\left( \mathrm{v}_{ph}-\mathrm{v}%
_{x}\right) +\left( \mathrm{v}_{ph}^{2}-c^{2}\right) \left( s\hbar k\right)
^{2}}{2\hbar \mathcal{E}}
\end{equation*}%
is the resonance width. The wave function (\ref{Psi1}) at the initial
condition $C_{s}\left( 0\right) =\delta _{s,0}$ describes inelastic
scattering of the electron on the slowed traveling wave. The electron energy
and momentum after the scattering are:%
\begin{equation*}
\mathcal{E}^{^{\prime }}=\mathcal{E}+s\hbar k\mathrm{v}_{ph},\ \
p_{x}^{^{\prime }}=p_{x}+s\hbar k,
\end{equation*}%
\begin{equation}
\ \mathbf{p}_{\bot }=\mathrm{const};\ \ s=0,\pm 1,\ldots .  \label{f1}
\end{equation}%
The probability of these processes are determined by the coefficients $%
C_{s}\left( t\right) :$ 
\begin{equation}
W_{s}=\left\vert C_{s}\left( t\right) \right\vert ^{2}.  \label{Prob}
\end{equation}

We will represent the solution of equations Eqs. (\ref{M2}) in the two
interaction regimes: diffraction --at the classical conditions of coherency (%
\ref{5.10}), (\ref{5.29}) and Bragg resonance --at the exact quantum
conditions of coherency with the quantum recoil of electron due to photon
absorption/radiation.

\subsection{Diffraction Regime of Electron Coherent Scattering on the
Traveling Wave Phase-Lattice}

For the finite electron-wave interaction time the interaction energy is
uncertain by the quantity $\delta \mathcal{E\simeq }\hbar /t_{int}$. The
diffraction regime of electron coherent scattering corresponds to the short
interaction times and intense wave fields. Thus, at the satisfaction of the
conditions 
\begin{equation*}
\delta \mathcal{E}\gg \hbar \left\vert \Gamma _{s}\right\vert ,\frac{%
\mathcal{W}_{eff}}{4\hbar \mathcal{E}}\gg \left\vert \Gamma _{s}\right\vert
\end{equation*}%
one can neglect the term $\sim \Gamma _{s}C_{s}\left( t\right) $ in Eqs. (%
\ref{M2}) and the dynamics of electron in the wavefields will be described
by the following equation \cite{LPB}: 
\begin{equation}
i\frac{\partial C_{s}\left( t\right) }{\partial t}=\frac{\mathcal{W}_{eff}}{%
4\hbar \mathcal{E}}\left( C_{s-1}\left( t\right) +C_{s+1}\left( t\right)
\right) .  \label{M3}
\end{equation}%
for which at the initial condition $C_{s}\left( 0\right) =\delta _{s,0}$ one
can obtain the following solution: 
\begin{equation*}
C_{s}\left( t\right) =J_{s}\left( \frac{1}{2\hbar \mathcal{E}}\int_{0}^{t}%
\mathcal{W}_{eff}dt^{\prime }\right) e^{-is\frac{\pi }{2}}.
\end{equation*}%
Consequently, the probability of this process according to Eq. (\ref{Prob})
will be done by the formula: 
\begin{equation}
W_{s}=J_{s}^{2}\left[ \frac{1}{2\hbar \mathcal{E}}\int_{0}^{t}\mathcal{W}%
_{eff}dt^{\prime }\right] .  \label{3.98}
\end{equation}%
In the case of a monochromatic wave from Eq. (\ref{3.98}) we have 
\begin{equation}
W_{s}=J_{s}^{2}\left( \frac{\mathcal{W}_{eff}t_{int}}{2\hbar \mathcal{E}}%
\right) ,  \label{Main}
\end{equation}%
where $t_{int}$ is the time-duration of the particle motion in the
wavefield. As is seen from Eq. (\ref{Main}) in the diffraction regime
symmetric diffraction ($J_{s}^{2}=J_{-s}^{2}$) into many momentum states is
possible. The process dynamics is defined by the argument of the Bessel
function $\alpha =\mathcal{W}_{eff}t_{int}/(2\hbar \mathcal{E})$. For $%
\alpha \lesssim 1$ only few diffraction maxima are possible. For the values $%
\alpha \gg 1$, the process is essentially multiphoton. The most probable
number of absorbed/emitted Cherenkov photons is $\overline{s}\simeq \alpha $%
. The width of the main diffraction maximums $\Delta (\overline{s})\simeq 
\overline{s}^{1/3}\hbar k$ and since $\overline{s}\gg 1$ then $\Delta (%
\overline{s})\ll \left\vert p_{x}^{^{\prime }}-p_{x}\right\vert $.

For the concreteness let us explicitly write probability for the Cherenkov
diffraction and Kapitza--Dirac effects. In case of Cherenkov diffraction of
electron on a traveling wave in a dielectric medium, from Eq. (\ref{Main})
we have 
\begin{equation}
W_{s}^{(\mathrm{Cherenkov})}=J_{s}^{2}\left( \xi \frac{mc^{2}}{\hbar }\frac{%
cp\sin \vartheta _{ch}}{\mathcal{E}}t_{int}\right) .  \label{3.99}
\end{equation}%
For the actual values of the parameters $\alpha \gg 1$, that is, the process
is essentially multiphoton. The most probable number of absorbed/emitted
Cherenkov photons is 
\begin{equation}
\overline{s}\simeq \xi \frac{mc^{2}}{\hbar }\frac{\mathrm{v}}{c}\sin
\vartheta _{ch}\cdot t_{int}.  \label{100'}
\end{equation}%
The scattering angles of the $s$-photon Cherenkov diffraction are determined
by the formula: 
\begin{equation}
\tan \vartheta _{s}=\frac{sn\hbar \omega \sin \vartheta _{ch}}{cp+sn\hbar
\omega \cos \vartheta _{ch}}.  \label{3.101}
\end{equation}%
From Eq. (\ref{3.101}) it follows that at the inelastic diffraction there is
an asymmetry in the angular distribution of the scattered particle: $%
\left\vert \vartheta _{-s}\right\vert >\vartheta _{s}$, i.e., the main
diffraction maxima are situated at different angles with respect to the
direction of particle initial motion. However, in accordance with the
condition $\left\vert s\right\vert n\hbar \omega /c\ll p$ of the eikonal
approximation this asymmetry is negligibly small and for the scattering
angles of the main diffraction maxima from Eq. (\ref{3.101}) we have $%
\vartheta _{-s}\simeq -\vartheta _{s}$. Hence, the main diffraction maxima
will be situated at the angles 
\begin{equation}
\vartheta _{\pm \overline{s}}=\pm \overline{s}\frac{n\hbar \omega }{cp}\sin
\vartheta _{ch}  \label{3.102}
\end{equation}%
with respect to the direction of the particle initial motion.

For the Kapitza--Dirac effect, which is the particular case of induced
Compton effect at $\omega _{1}=\omega _{2}\equiv \omega $, $E_{1}=E_{2}=E$,
from Eqs. (\ref{Psi1}) and (\ref{Main}) we can write%
\begin{equation}
\Psi =e^{-\frac{i}{\hbar }\mathcal{E}t}e^{\frac{i}{\hbar }\mathbf{p}_{\bot }%
\mathbf{r}}\sum_{s}e^{-is\frac{\pi }{2}}J_{s}\left( \frac{%
e^{2}c^{2}E^{2}t_{int}}{2\mathcal{E}\hbar \omega ^{2}}\right) e^{\frac{i}{%
\hbar }\left( p_{x}+2s\hbar \frac{\omega }{c}\right) x}.  \label{KP}
\end{equation}%
Hence, the probability of $s$-photons Kapitza--Dirac diffraction effect on
the strong standing EM wave in vacuum is determined by the formula: 
\begin{equation}
W_{s}^{(Kapitza-Dirac)}=J_{s}^{2}\left( \frac{e^{2}c^{2}E^{2}t_{int}}{2%
\mathcal{E}\hbar \omega ^{2}}\right) .  \label{KP2}
\end{equation}%
Note that formula (\ref{KP2}) for the nonrelativistic case $\mathcal{E}%
\simeq mc^{2}$ coincides with analogous formula (7) of the paper \cite{Bat}
up to a factor of $1/2$ (in the paper \cite{Bat}) the factor $1/2$ has been
missed perhaps at the transformation of recurrent relation of Bessel
functions $J_{s}\left( x\right) $).

In general case of inelastic Compton diffraction on counterpropagating waves
of different frequencies and electron diffraction on a travelling EM wave in
an undulator/wiggler the probabilities of $s$-photon diffraction effect may
be written by analogy of formulas (\ref{3.99}) and (\ref{KP2}) with
corresponding parameters.

Note that the general formula (\ref{3.99}) for Cherenkov diffraction that
expresses the probabilities of multiphoton absorption/radiation processes by
electron at the induced Cherenkov effect has been applied in the paper \cite%
{ushir} for explanation of the experiment on energetic widening of an
electron beam at the induced Cherenkov process in a gaseous medium,
implemented in SLAC \cite{P1} (see, also the next experiment of this group 
\cite{P3}, made in the same conditions).

Formula (\ref{3.99}) for Cherenkov diffraction in a dielectric medium, which
had been received forty years ago \cite{dif.3} and included in the
monographs \cite{2006} and \cite{2016}, recently has been considered in the
paper \cite{LPB} because of reproduction of electron diffraction effect on a
travelling EM wave in a dielectric medium by the authors of the paper \cite%
{mis}, nevertheless represented it as a new phenomenon predicted in the
paper \cite{mis}. Repeating the results of the paper \cite{dif.3} with the
main formula (\ref{3.99}) in other way of derivation -within the
Helmholtz--Kirchhoff diffraction theory (as was mentioned in Introduction,
by the method developed in the paper \cite{Echl-Lub}), authors report in the
Abstract that they showed on the possibility of electrons diffraction effect
on a travelling EM wave in a dielectric medium. Apart from such
misunderstanding resulted to serious confusion in scientific literature, it
is interesting the claim of the authors even in the Abstract of the paper 
\cite{mis} (and this is the only difference from the paper \cite{dif.3})
that explains diffraction effect by the "group velocity of light" (and this
is repeated also in the text -and for a monochromatic wave!). This is
interesting fact since repeating the results, at the same time authors of
the paper \cite{mis} quote to original papers \cite{dif.2,dif.3,dif.4}
without a comment. While, as it has been shown above, diffraction effect is
thoroughly the result of the phase relations and is conditioned
exceptionally by the phase velocity of light that must be smaller than $c$
-- just the physical reason which causes electron diffraction on a light
phase lattice due to phase matching between the particle and wave phase
velocity. Beside this rough mistake, authors of the paper \cite{mis} ignored
the existence of critical field in this process and influence of considered
phenomenon of particle "reflection" or capture on the diffraction effect.
Meanwhile, one of the authors \cite{mis} is also a co-author of both
inelastic diffraction effect \cite{dif.4} and "reflection" phenomenon in the
undulators \cite{Und._Reflec}.

\subsection{Bragg Regime of Electron Coherent Scattering on the Traveling
Wave Phase-Lattice}

For the diffraction effect with sufficiently long interaction time one can
fulfill resonance condition with quantum recoil for the concrete $s_{0}$: $%
\Gamma _{s_{0}}=0$. The latter can be written as 
\begin{equation}
\mathrm{v}_{ph}-\mathrm{v}_{x}=\left( c^{2}-\mathrm{v}_{ph}^{2}\right) \frac{%
s_{0}\hbar k}{2\mathcal{E}}.  \label{res_q}
\end{equation}

The condition (\ref{res_q}) has transparent physical interpretation in the
intrinsic frame of reference of the slowed wave. In this frame, due to the
conservation of particle energy and transverse momentum the real transitions
in this strongly quantum regime occur from a $p_{x}^{\prime }$ state to the $%
-p_{x}^{\prime }$ one and we reach the Bragg diffraction effect on a slowed
traveling wave at the fulfilment of the condition: 
\begin{equation}
2p_{x}^{\prime }=-s_{0}\hbar k^{\prime }\qquad (s_{0}=\pm 1;\pm 2...).
\label{3.20}
\end{equation}%
The latter expresses the condition of exact resonance between the particle
de Broglie wave and the \textquotedblleft wave motionless
lattice\textquotedblright . In particular, in this case when the above
mentioned particle capture regime by the slowed traveling wave \cite{2006}
takes place, we have the quantum effect of zone structure of particle states
like the particle states in a crystal lattice, and at the condition (\ref%
{res_q}) the diffraction maxima take place see Fig. 1. Here we just write
the solution for the resonant case $\Gamma _{1}=0.$At the condition%
\begin{equation}
\delta \mathcal{E}\ll \hbar \left\vert \Gamma _{s}\right\vert ,\frac{%
\mathcal{W}_{eff}}{4\hbar \mathcal{E}}\ll \left\vert \Gamma _{s}\right\vert
;s\neq 0,1  \label{BCOn}
\end{equation}%
from the set of equations (\ref{M2}) one can keep only resonant ones for $%
C_{0}$ and $C_{1}$:%
\begin{equation*}
i\frac{\partial C_{01}\left( t\right) }{\partial t}=\frac{\mathcal{W}_{eff}}{%
4\hbar \mathcal{E}}C_{10}\left( t\right) ,
\end{equation*}%
with the solution%
\begin{eqnarray}
C_{0}\left( t\right) &=&\cos \left( \frac{1}{4\hbar \mathcal{E}}\int_{0}^{t}%
\mathcal{W}_{eff}dt^{\prime }\right)  \notag \\
C_{1}\left( t\right) &=&-i\sin \left( \frac{1}{4\hbar \mathcal{E}}%
\int_{0}^{t}\mathcal{W}_{eff}dt^{\prime }\right)  \label{solb}
\end{eqnarray}%
For the concreteness let us explicitly write probability for Kapitza--Dirac
effect - $\omega _{1}=\omega _{2}\equiv \omega $, $E_{1}=E_{2}=E_{0}$, $%
\mathrm{v}_{ph}=0,k=2\omega /c$). From Eq. (\ref{res_q}) at $s_{0}=1$ we
obtain resonant initial momentum $p_{x}=-\hbar \frac{\omega }{c}$ and wave
function can be writen as 
\begin{equation}
\Psi =e^{\frac{i}{\hbar }\mathbf{p}_{\bot }\mathbf{r}-\frac{i}{\hbar }%
\mathcal{E}t}\left[ C_{0}\left( t\right) e^{-i\frac{\omega }{c}%
x}+C_{1}\left( t\right) e^{i\frac{\omega }{c}x}\right] .  \label{w}
\end{equation}%
with the probabilities 
\begin{eqnarray}
W_{0} &=&\cos ^{2}\left( \frac{e^{2}c^{2}E_{0}^{2}t_{int}}{4\mathcal{E}\hbar
\omega ^{2}}\right) ,  \notag \\
W_{1} &=&\sin ^{2}\left( \frac{e^{2}c^{2}E_{0}^{2}t_{int}}{4\mathcal{E}\hbar
\omega ^{2}}\right) .  \label{BregP}
\end{eqnarray}%
For the nonrelativistic case $\mathcal{E}\simeq mc^{2}$ Eqs. (\ref{BregP})
coinside with the results in Ref. \cite{Bat}.

This work was supported by SCS of RA.

\end{document}